\documentclass[pra,aps,10pt,superscriptaddress,preprint,floatfix]{revtex4-2}
\usepackage{graphicx,color,xcolor}
\usepackage[nice]{nicefrac}	
\usepackage{amsmath,amssymb,bm}
\usepackage{amsmath}
\usepackage{braket}
\usepackage[rightcaption]{sidecap}
\usepackage{lineno}
\usepackage{etoolbox}
\makeatletter
\patchcmd{\frontmatter@abstract@produce}
  {\vskip200\p@\@plus1fil
   \penalty-200\relax
   \vskip-200\p@\@plus-1fil}
  {}
  {}
  {}
\makeatother

\usepackage[plainpages=false,pdfpagelabels,colorlinks=true,linkcolor=red,urlcolor=blue,citecolor=blue,pdftitle={},pdfauthor={},pdfdisplaydoctitle=true,pdfduplex=DuplexFlipLongEdge]{hyperref}

\definecolor{darkred}{rgb}{0.90,0.2,0.2}
\definecolor{darkgreen}{rgb}{0,0.60,.2}
\definecolor{darkblue}{rgb}{0.1,0.3,1}
\definecolor{grey}{cmyk}{0,0,0,0.25}
\definecolor{orange}{cmyk}{0,0.6,0.8,0}

\usepackage{multibib}
\newcites{M}{Methods References}
\newcites{SI}{Supplemental Information References}

\begin{document}


%
%

\title{Direct observation of hydrodynamization and local prethermalization}

\author{Yuan Le}
\affiliation{Department of Physics, The Pennsylvania State University, University Park, Pennsylvania 16802, USA}
\author{Yicheng Zhang}
\affiliation{Department of Physics, The Pennsylvania State University, University Park, Pennsylvania 16802, USA}
\author{Sarang Gopalakrishnan}
\affiliation{Department of Physics, The Pennsylvania State University, University Park, Pennsylvania 16802, USA}
\affiliation{Department of Electrical and Computer Engineering, Princeton University, Princeton, New Jersey 08544, USA}
\author{Marcos Rigol}
\affiliation{Department of Physics, The Pennsylvania State University, University Park, Pennsylvania 16802, USA}
\author{David S. Weiss}
\affiliation{Department of Physics, The Pennsylvania State University, University Park, Pennsylvania 16802, USA}

\maketitle


\vspace*{1.5cm}

{\bf Hydrodynamics accurately describes relativistic heavy-ion collision experiments well before local thermal equilibrium is established~\cite{Rapp_2004}. This unexpectedly rapid onset of hydrodynamics---which takes place on the fastest available timescale---is called hydrodynamization~\cite{Florkowski_2018, Schenke_2021, RevModPhys_2021}. It occurs when an interacting quantum system is quenched with an energy density that is much greater than its initial energy density~\cite{prethermalization, Borsanyi}. During hydrodynamization, energy gets redistributed across very different energy scales. Hydrodynamization precedes local equilibration among momentum modes~\cite{prethermalization}, which is local prethermalization to a generalized Gibbs ensemble~\cite{rigol_dunjko_07, schmiedmayer_prethermalization} in nearly integrable systems or local thermalization in non-integrable systems~\cite{D_Alessio_2016}. Many theories of quantum dynamics postulate local (pre)thermalization~\cite{mallayya_rigol_19, doyon2019lecture}, but the associated timescale has not been quantitatively studied. Here we use an array of 1D Bose gases to directly observe both hydrodynamization and local prethermalization. After we apply a Bragg scattering pulse, hydrodynamization is evident in the fast redistribution of energy among distant momentum modes, which occurs on timescales associated with the Bragg peak energies. Local prethermalization can be seen in the slower redistribution of occupation among nearby momentum modes. We find that the time scale for local prethermalization in our system is inversely proportional to the momenta involved. During hydrodynamization and local prethermalization, existing theories cannot quantitatively model our experiment. Exact theoretical calculations in the Tonks-Girardeau limit~\cite{girardeau1960relationship} show qualitatively similar features.}

\newpage

Hydrodynamization has been theoretically explored in high energy~\cite{Florkowski_2018, prethermalization,Schenke_2021, RevModPhys_2021, Spalinksy_2015, schlichting_2019}, cosmological~\cite{mottola_1997}, and photo-excited correlated metal~\cite{Kollar} contexts. These calculations typically involve simplified Hamiltonians that do not directly correspond to measurable physical systems. The process of hydrodynamization has not been directly observed in any system. We observe hydrodynamization using trapped 1D Bose gases with point contact interactions (Lieb-Lininger (LL) gases~\cite{lieb1963exact}), quenching them with a Bragg pulse. The high energy density of the pulse compared to the initial energy density of our ultracold gases allows us to realize this universal phenomenon. The near-integrability of trapped LL gases provides a framework from which we draw a general picture that applies to non-integrable systems. 

Since the strong coupling Tonks-Girardeau (TG) limit of our 1D Bose gases is exactly solvable, we first explain the physics of hydrodynamization and local prethermalization for a homogeneous TG gas. As we will see, our theoretical insights transfer to trapped systems with finite-coupling like the ones in the experiments. The TG Hamiltonian can be mapped onto noninteracting fermions by a change of variables that acts nonlocally on the boson field operators. The pre-quench ground state is a Fermi sea, with a many-body wavefunction given by $\prod_{\mid\theta \mid <\theta_F} \hat c^\dagger_{\theta}|0\rangle$, where $\hat c^\dagger_\theta$ creates a fermion with momentum $\theta$, and $\theta_F$ is the Fermi momentum. Anticipating generalization to the finite-coupling regime, we will refer to the fermions as quasiparticles, and their momenta as rapidities. 

\sloppy A Bragg pulse in the Kapitza-Dirac limit transforms the initial wavefunction to $\prod_{\mid\theta \mid<\theta_F} (\sum_n J_n(a) e^{-it\hbar (\theta+n2k)^2/2m} \hat c^\dagger_{\theta+n2k}) |0\rangle$, where $k=2\pi/\lambda$, $J_i(a)$ are Bessel functions of the first kind, $n$ is an integer, and $a$ depends on the Bragg pulse area~\cite{Cauxseptimescales_2016}. The rapidity distribution is conserved after the pulse and does not evolve, although the many-body wavefunction evolves rapidly due to the spread of phase factors associated with each quasiparticle. Since momentum modes of the bare particles are complicated superpositions of quasiparticle states, relative phase evolution in the rapidity basis results in a change of occupancy of momentum modes, which evolve at a rate associated with the highest energy difference between rapidity states. For $a\lesssim1$, the rate is dominated by the energy difference between the $n$=0 and $\pm 1$ Bragg orders, which is $(2\hbar k)^2/2m$, where $m$ is the atomic mass.

Since it happens so fast, the long-lived nature of the quasiparticles and the conservation of rapidities associated with integrability are not necessary for hydrodynamization. Non-integrable interacting many-body systems allow for a similar quasiparticle description on some short time scale. As long as the quench energy significantly exceeds the initial energy density, one expects that these quasiparticles will live much longer than the hydrodynamization time scale. So the above description of hydrodynamization applies.

The Bragg pulse quench provides a natural spatial view of hydrodynamization. It creates a $\lambda/2$ scale density oscillation, which tends to flatten on the time scale that the $|n|=1$ side momentum peaks particles traverse $\lambda/2$, $(\lambda/2)/(2\hbar k/m)$, i.e., on the order of $2\pi/\omega_{hd}$, where $\omega_{hd}=\hbar(2k)^2/m$. Viewed another way, the Bragg pulse creates two overlapped, oppositely-directed, rapidly moving clouds that collide with each other and the stationary cloud. Hydrodynamization can occur even before the center of masses of these clouds have traversed the average interparticle spacing. The collision energy, imparted with the creation of the Bragg peaks, gets partially transferred to the momentum modes between the moving side peaks on hydrodynamization time scales. The situation is qualitatively similar in relativistic heavy ion collisions, where part of the kinetic energy in the center of mass frame is rapidly redistributed from the colliding nuclei to the stationary emerging quark-gluon plasma. Quantitatively, the scale of collision energies in the two systems, which is inversely proportional to the timescale of hydrodynamization, differs by 18 orders of magnitude. 

After hydrodynamization, the correlations in the many-body wavefunction have not yet locally equilibrated to the post-quench condition. We call this subsequent evolution local prethermalization. We can model this process in the TG limit as follows: the occupation of momentum mode $p$ is $f(p)\equiv \langle \hat b^\dagger_p \hat b^{}_p \rangle = \int dx dy \exp[ip(x-y)/\hbar]\langle \hat c^\dagger_x \hat S^{}_{x-y} \hat c^{}_{y}\rangle$, where $\hat S_{x-y}$ is a ``string operator'' that counts quasiparticle number fluctuations between positions $x$ and $y$~\cite{minguzzi_gangardt_05, del2022hydrodynamic}.
$f(p)$ is dominated by the nonlocal fermionic real-space correlation function $\langle \hat c^\dagger_x \hat S^{}_{2\hbar \pi/p} \hat c^{}_{x+2\hbar \pi/p}\rangle$, with the initial zero-temperature state having weaker particle fluctuations than the locally equilibrated final state. Fluctuations grow after the quench, as quasiparticles from the side peaks traverse the distance $2\hbar \pi/p$. The time it takes for these fluctuations (and therefore the string operator $\hat S_{2\hbar \pi/p}$) to equilibrate for momentum $p$ is $\tau_p \sim 2\hbar \pi / (p v_{\pm1})$, where $v_{\pm1}=2\hbar k/m$ is the group velocity of the side peaks. Thus, the local prethermalization of the bosonic momentum distribution involves a spectrum of timescales. That the bare-particle momentum distribution equilibrates on a timescale set by the spreading of excited quasiparticles holds true in general for nearly integrable 1D systems with long-lived quasi-particles, as in our experimental setup~\cite{wilson_malvania_20, Malvania_2021}. Using the local density approximation, it also applies to trapped systems as long as the density profile does not change significantly during local prethermalization.

\begin{figure}[!t]
\includegraphics[width=0.9\columnwidth]{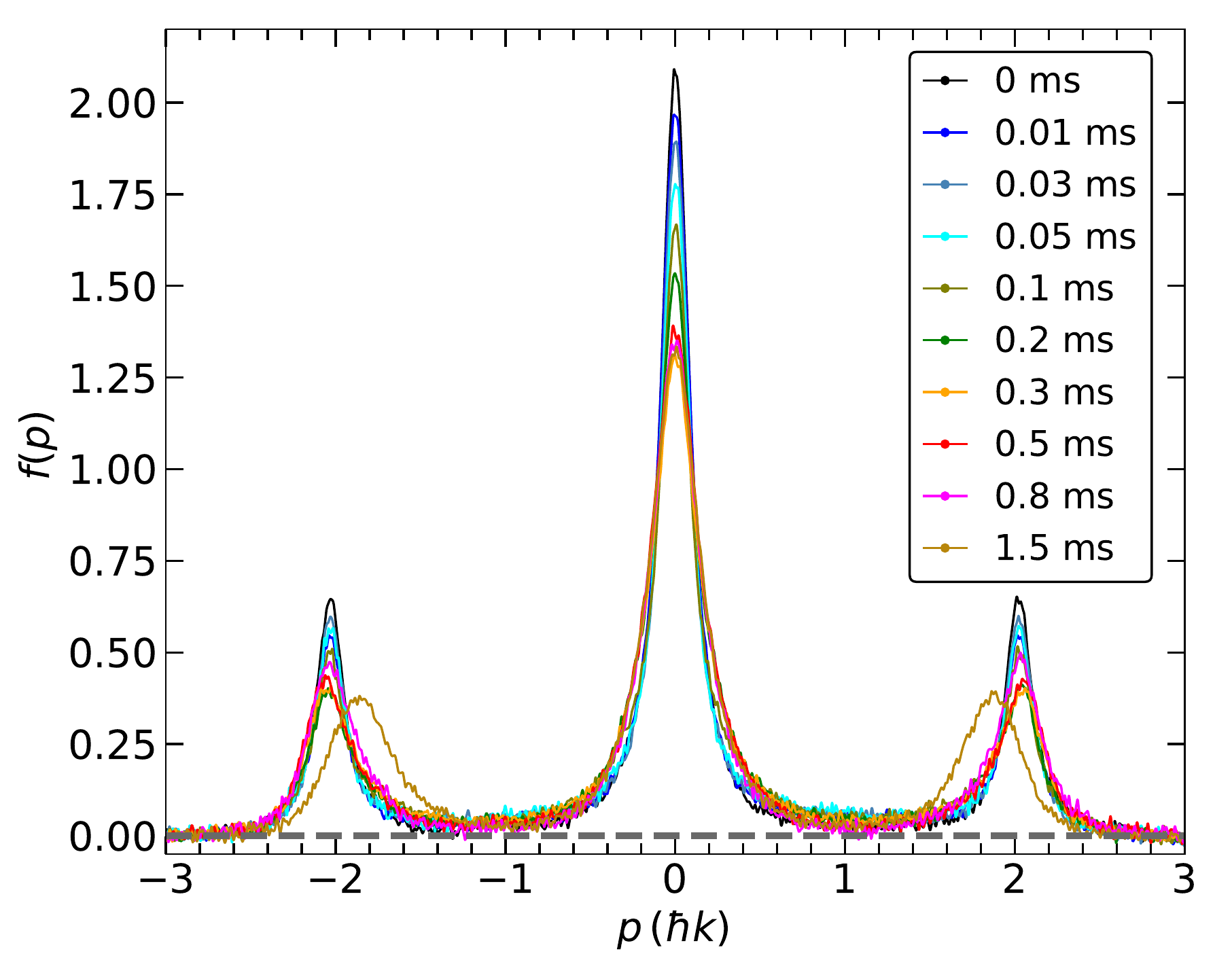}
\caption{\textbf{Evolving momentum distributions.} Each curve is the momentum distribution for $\bar{\gamma}_0=2.3$ at some time after the Bragg scattering quench. Before 0.5 ms, the side peaks have not yet noticeably been affected by the trap so they remain at their initial momenta.}
\label{momentum}
\end{figure}

Our experiment starts with a bundle of nearly zero temperature 1D gases consisting of $^{87}$Rb atoms confined in a blue-detuned 2D lattice, with axial trapping provided by crossed red-detuned dipole trapping beams (see Methods). We pulse on an axial lattice beam with wavevector $k_0=2\pi$/(775 nm) for 6 $\mu$s and then measure the momentum distribution (see Methods) as a function of the time $t_{ev}$ after the pulse. Fig.~\ref{momentum} shows a set of momentum distributions for an initial (pre-Bragg pulse) weighted average dimensionless interaction strength, $\bar{\gamma}_0$, of 2.3 (see Methods). The curves range over the first 1.5 ms, which is short compared to the axial oscillation period of 16.7 ms. The expected Bragg side-peaks are visible, as is their eventual slowing down (see the 1.5 ms curve) due to the axial trap. In what follows we focus on short times, for which the side peaks are not significantly affected by the presence of the trap and remain at their initial momenta. Corresponding TG theory curves are shown in Extended Data Fig.~1. 

\begin{figure}
\includegraphics[width=0.85\columnwidth]{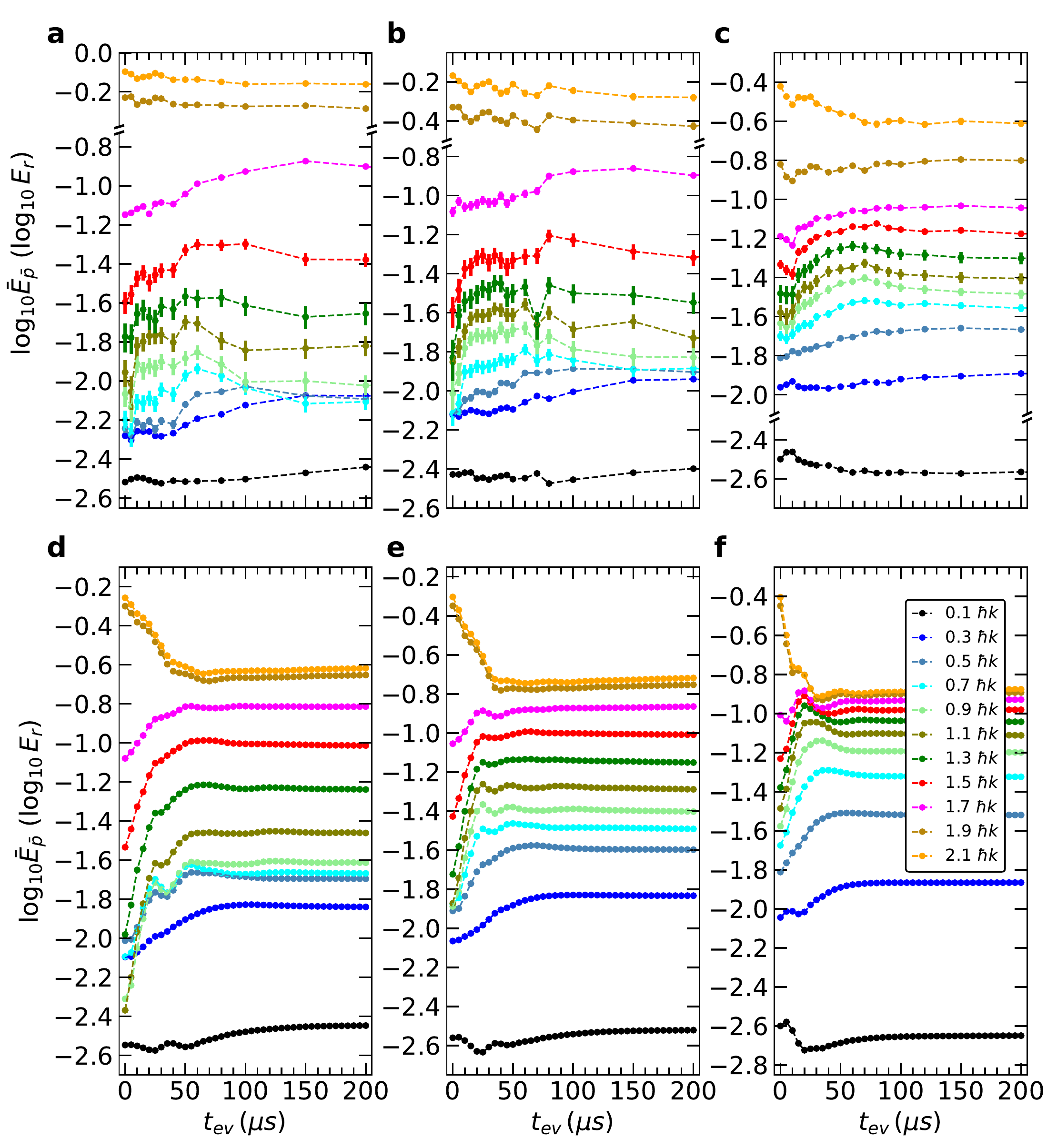}\hspace{0.1cm}
\caption{\textbf{Hydrodynamization.} Each curve is the time evolution of the integrated energy in a different 0.2$\hbar k_0$ wide momentum group. The central momentum of each color-coded group is shown in the legend. \textbf{a--c} Experimental curves, for $\bar{\gamma}_0=3.4$, 2.3, and 0.94, respectively. See Methods for an explanation of the error bars. \textbf{d--f} Theoretical curves for single 1D gases in the TG-gas limit. The average energy per particle of the three theory figures are the same as the experimental figures immediately above them. Hydrodynamization is evident in the $\sim 33$ $\mu$s period oscillations of the few lowest and highest momentum groups in each figure. The most dramatic feature of hydrodynamization is the very rapid initial energy changes in the intermediate momentum groups. Variations among the different experimental and theoretical figures are discussed in the text. The dashed lines are guides to the eye.}
\label{hydrodynamization}
\end{figure}

Hydrodynamization causes rapid changes in the energy distribution of the system. To see it, we integrate the kinetic energy in successive momentum ranges (denoted by their average momentum, $\bar{p}$) and plot those integrated energies as a function of time. Experimental data for momentum groups up to the first Bragg peak are shown in Fig.~\ref{hydrodynamization}a--c for $\bar{\gamma}_0=$ 3.4, 2.3, and 0.94, respectively. Results from TG-gas theory are shown in Fig.~\ref{hydrodynamization}d--f, for single 1D gases with the same average energy per particle as the experimental curves directly above them. The early time dynamics are due to hydrodynamization. Extended Data Figs.~2 and 3 respectively show the evolution of the experimental rapidity distribution for $\bar{\gamma}_0=2.3$, and the theoretical TG rapidity distribution with the same average rapidity energy (see also Ref.~\cite{Cauxseptimescales_2016}). Extended Data Fig.~4 shows the rapidities version of Fig.~\ref{hydrodynamization}b. As expected, the rapidity distribution does not evolve on these timescales.

The details of Fig.~\ref{hydrodynamization} cannot be predicted without a dynamical model that can be solved at finite coupling after this strong quench. From the experimental results, several general observations can be made. Fig.~\ref{hydrodynamization}b is the clearest experimental result to interpret, as it is at a ``sweet spot'' where the three dominant momentum peaks are well-resolved but the occupation of the intermediate momentum modes is not too small. The $\bar{p}=$ 1.9 and 2.1 curves show an oscillation period of $T_{hd}\sim2\pi/\omega_{hd}=33\mu s$, and the $\bar{p}=$ 0.1, 0.3, and 0.5 curves show an out of phase oscillation. The intermediate curves from $\bar{p}=$ 0.7 to 1.5 all show a rapid rise with a characteristic time of $\sim 2/\omega_{hd}$, the result of rapid dephasing among the many rapidity modes that contribute to these momentum modes.

Fig.~\ref{hydrodynamization}a has qualitatively similar hydrodynamization features to Fig.~\ref{hydrodynamization}b. The $\omega_{hd}$ oscillations of the side momentum peak ($\bar{p}=$ 1.9 and 2.1) are less pronounced than in Fig.~\ref{hydrodynamization}b, suggesting that $\pm2\hbar k_0$ momenta are less composed of the central peak rapidities. The intermediate $\bar{p}$ curves, which must still have similar rapidity components as in Fig.~\ref{hydrodynamization}b, also rise, but with more oscillations at $\omega_{hd}$, which seems reasonable given the narrower rapidity distributions and correspondingly slower dephasing times. The dephasing in Fig.~\ref{hydrodynamization}c is faster in all momentum curves, a consequence of wider rapidity distributions. In the intermediate $\bar{p}$ curves, hydrodynamization manifests as rapid initial downturns. 

The various hydrodynamization features of the TG gas theory shown in Fig.~\ref{hydrodynamization}d--f are similar to those in the associated experiments. One would not expect them to be identical because the mapping between rapidities and momenta are different for different values of $\gamma$. The hydrodynamization timescale is predictably the same for all these results, and the hydrodynamization dephasing times are qualitatively similar for experiment and theory curves with the same rapidity energy per particle (those in the same column). The biggest difference between the theory and experiment (other than the absence of noise in the theory) is that the energy changes during hydrodynamization are about 2.5 times greater in the theory. One can perhaps qualitatively understand this because the pre-quench kinetic energy in our experimental range of $\bar{\gamma}_0$ ranges from 20\% to 65\% of the rapidity energy, with the rest being interaction energy. For infinite $\gamma$, all the energy is kinetic.

The onset of local prethermalization is also visible in Fig.~\ref{hydrodynamization}. After hydrodynamization dephasing is complete, which takes less than $\sim 2T_{hd}$ ($\sim 60 \mu s$) for all curves, the energies continue to evolve, approaching their GGE values on progressively longer timescales as the momentum decreases. This can be most clearly seen in the theory (Figs.~\ref{hydrodynamization}d--f) and in the lower $\bar\gamma_0$ experimental results (Fig.~\ref{hydrodynamization}c), looking from the $\bar{p}$=1.5 to 0.3 curves in each. Local prethermalization is more than twice as fast in the theory than in the experiments with the same rapidity energy. It is, however, difficult to consistently extract local prethermalization time constants because of the diversity of curve shapes and the modest separation of hydrodynamization and local prethermalization time scales in this momentum range. 

\begin{figure}[!t]
\includegraphics[width=0.6\columnwidth]{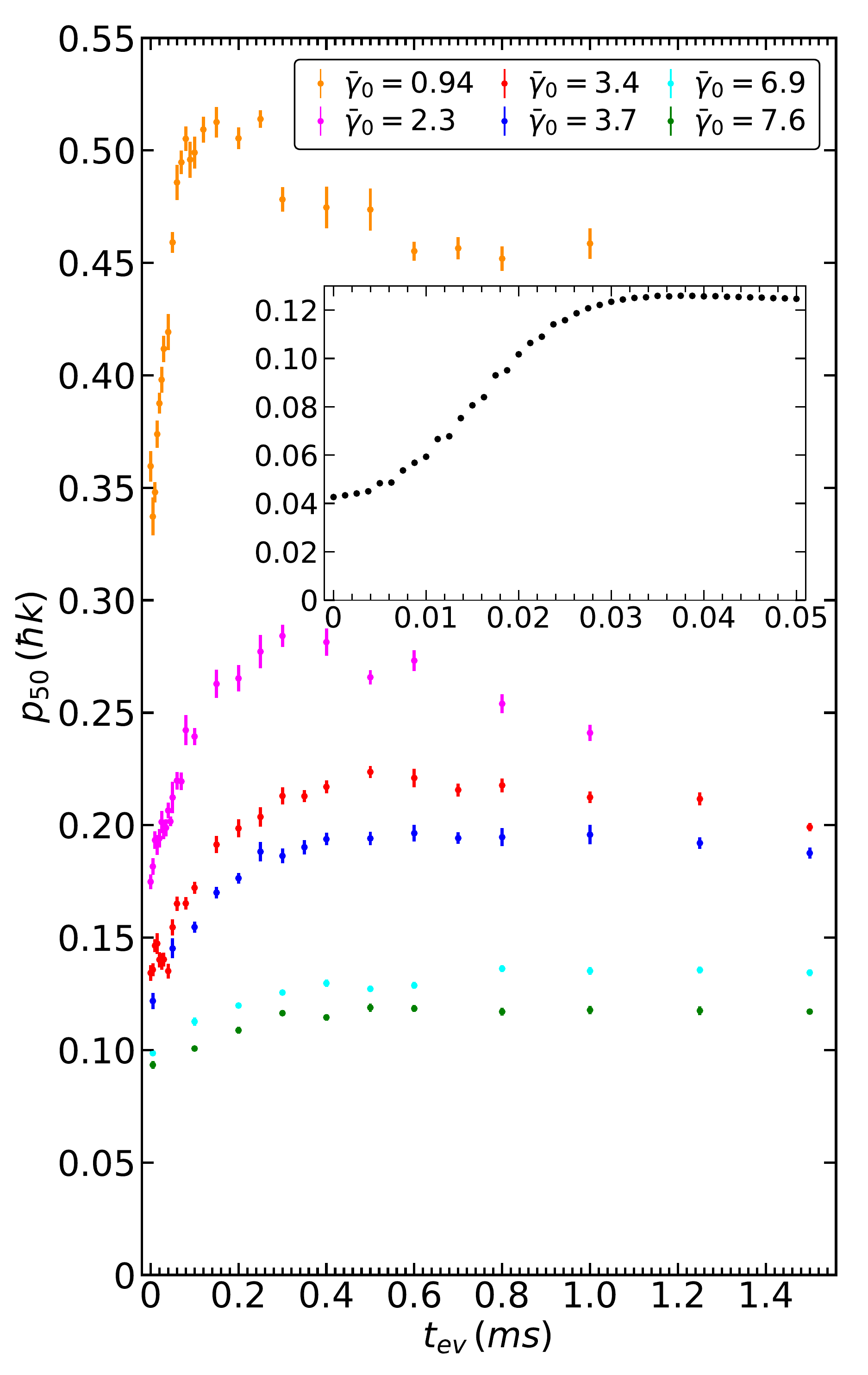}
\caption{\textbf{Local prethermalization.} For a range of $\bar\gamma_0$ we plot $p_{50}$, the momentum line dividing the atoms in the central peak in half, as a function of the time after the Bragg pulse. The different colored curves correspond to different experimental $\bar{\gamma}_0$, as shown in the key. The inset shows a typical TG theory $p_{50}(t_{ev})$ curve, in this case for a single 1D gas with $k=4k_0$ and a peak 1D density of 1.36~($\mu \text{m})^{-1}$. Small hydrodynamization oscillations are just barely visible throughout local prethermalization in the theory, and at the earliest times in the experiment.}
\label{prethermalization}
\end{figure}

To better study local prethermalization, we focus on the redistribution among momentum modes where the occupations are large and vary rapidly with momentum. We find a robust observable in the momentum dividing line, $p_f$, between the $f\%$ of lower momentum atoms in the central peak (between $\pm\hbar k_0$) and the rest. $f=50$ selects a momentum near the full width at half maximum (FWHM) of the central peak. Fig.~\ref{prethermalization} shows a set of experimental $p_{50}$ curves as a function of time (see Extended Data Figs.~5 and 6 for curves that show other possible observables). We extract $\tau_{p_{50}}$ from each of the curves in Fig.~\ref{prethermalization} by finding the time at which $p_{50}$ reaches half its peak value (see Methods).

\begin{figure}[!t]
\includegraphics[width=0.7\columnwidth]{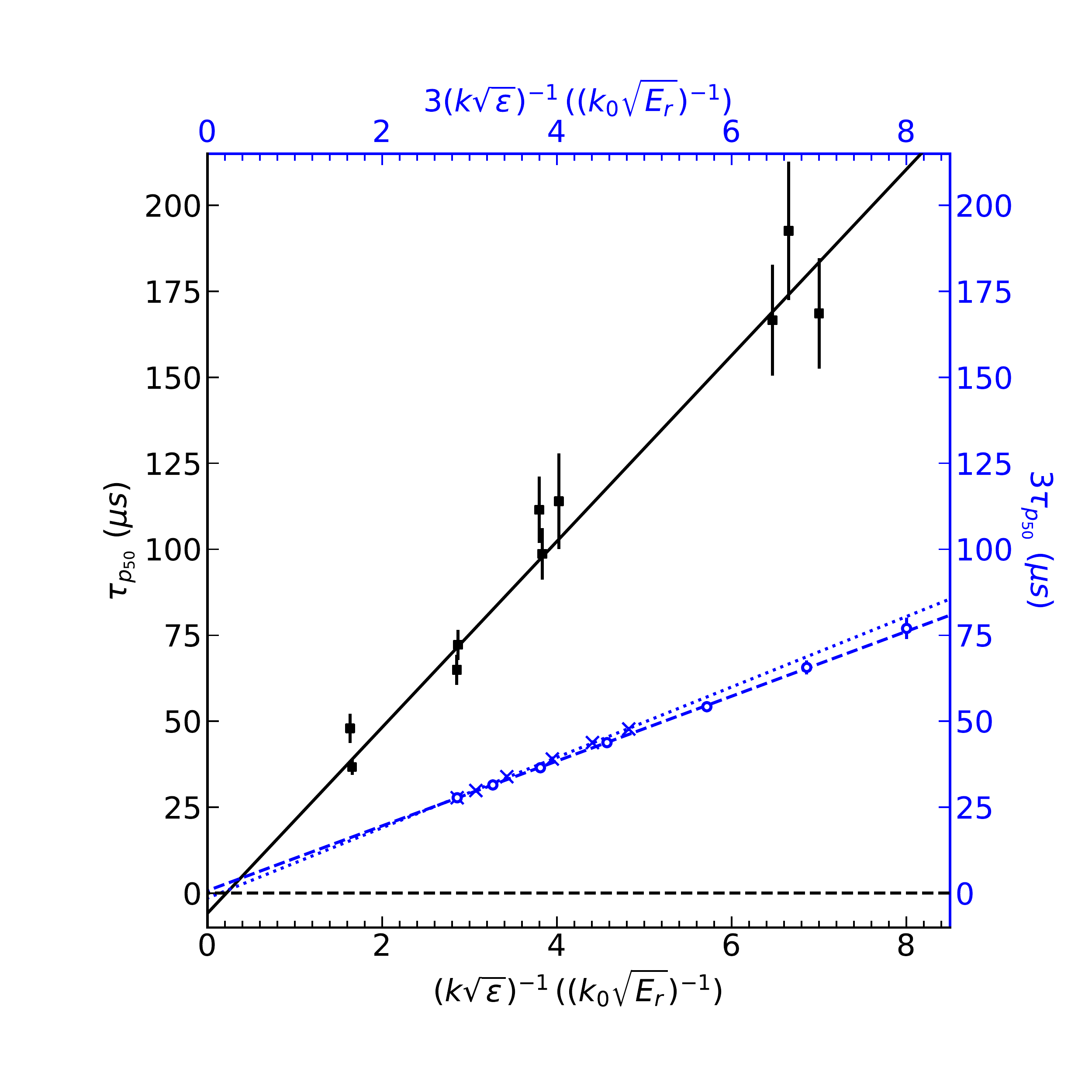}
\caption{\textbf{Local prethermalization time scales vs $(k\sqrt{\epsilon})^{-1}$.} The black experimental points show the time constants extracted from curves like those in Fig.~\ref{prethermalization}. See Methods for an explanation of the error bars. The blue points use the upper and right axes, and show the time constants extracted from curves like those in the inset of Fig.~\ref{prethermalization} for a fixed $k=4k_0$ (exes) or a fixed $\epsilon=0.06885\hbar k$ (open circles). The solid, dotted and dashed lines are least squares fits to the associated points; all the intercepts are zero to within a few standard deviations ($-5.8 \pm 4.9$, $-0.47 \pm 0.13$, $0.25 \pm 0.081$ for the black squares, blue exes, and blue circles, respectively). These results validate our qualitative model of the local prethermalization rate scaling inversely with $p$ and the Bragg peak velocity. The slope of the experimental line is 2.6 times the slope of the theory lines. }
\label{scaling}
\end{figure}

In Fig.~\ref{scaling}, we plot $\tau_{p_{50}}$ for a range of $\bar{\gamma_0}$, choosing the abscissa to be $1/\sqrt{\epsilon_0}$, where $\epsilon_0$ is the average pre-quench energy per particle from LL theory. The data fits to a straight line with an intercept close to zero. We also find that $p_{50}$ at the midpoint of local prethermalization, $p^m_{50}$, is proportional to $\sqrt{\epsilon_0}$ (see Extended Data Fig. 7). Qualitatively, we expect $\tau_{p_{50}}$ to be inversely proportional to the effective momentum $\bar{p}_{50}$ being measured. The linear fits in Fig.~\ref{scaling} and Extended Data Fig.~7 together strongly imply that both $\sqrt{\epsilon_0}$ and $p^m_{50}$ are proportional to $\bar{p}_{50}$, so that the local prethermalization rate is in fact inversely proportional to momentum. The linear fits also work well for $f$ = 40 and 60 (see Extended Data Fig.~7a and c), although the ratio of $\bar{p}_f$ to $p^m_{f}$ depends on $f$.

Because the hydrodynamization features in the theory are larger and local prethermalization is faster, we calculate $p_{50}$ for simulations with $k=4k_0$ in order to achieve a better separation of time scales ($1/\omega_{hd}\propto1/k^2$ while $\tau_{p_{50}}\propto 1/k)$. An example of $p_{50}(t_{ev})$ is shown in the inset to Fig.~\ref{prethermalization}, and the theoretical results for $\tau_{p_{50}}$ vs $1/\sqrt{\epsilon}$ are the blue exes in Fig.~\ref{scaling}. The linear fit crosses near zero, as in the experiment. The blue open circles show $\tau_{p_{50}}$ vs $k$. That its linear fit also crosses near zero validates the velocity part of our dimensional argument. While both the experiment and theory show that $\tau_{p_{50}}= C/(k\sqrt{\epsilon})$, where $C$ is a constant and $C_\text{exp} = 2.6C_\text{th}$, which is consistent with the results in Fig.~\ref{hydrodynamization}. The slower local prethermalization in the experiment is presumably related to its (fixed) finite-strength $\delta$-function interactions. (It is neither due to temperature effects nor to the average over 1D gases (see Extended Data Fig. 8)). The time for the TG gas to reach the local GGE is about twice $\tau_{p_{50}}$ (see Fig.~\ref{prethermalization} inset), which we find is $\sim5$ times faster than the time it takes the Bragg scattered components to traverse the wavelength associated with $p^m_{50}$. 

We have used nearly integrable 1D Bose gases to elucidate the behavior of many-body quantum systems immediately after a rapid, high energy quench. There are two distinct time scales. The first is hydrodynamization, during which short distance variations in the wavefunctions smooth out and energy is rapidly redistributed among distant momentum modes. Although hydrodynamization is qualitatively the same in a wide range of many-body quantum systems, it is easier to study with cold atoms because one can rapidly shut-off interactions and thus measure evolving momentum distributions with good time resolution. The second is local prethermalization, which in our experiment involves a redistribution among nearby momentum modes on timescales that vary inversely with momentum. We find that local prethermalization is surprisingly fast. For nearly integrable systems like the trapped LL gas, the system approaches the local GGE during local prethermalization. Local prethermalization can also occur in far from integrable systems when there are extra conserved quantities~\cite{mallayya_rigol_19}. 

Outside of the hard-core limit, no existing theory can calculate the complete dynamics during hydrodynamization and local prethermalization, which makes it an important frontier of many-body dynamical theory. Generalized hydrodynamics (GHD) was developed to calculate the dynamics of the rapidity distribution after local prethermalization~\cite{castro2016emergent, bertini2016transport, schemmer2019generalized, Malvania_2021}. We suspect, however, that GHD still works during local prethermalization, i.e., that it describes the evolution of the rapidity distributions in space even though the local GGE is not yet satisfied. The situation would be analogous to the quark-gluon plasma, where hydrodynamics works after hydrodynamization but before local thermalization.  If confirmed, this will imply a further relaxation of the required conditions for GHD~\cite{Malvania_2021}, one that can be tested theoretically in the hard-core limit and experimentally with ultracold atoms.

\acknowledgements
We acknowledge Neel Malvania for preliminary experimental work related to this paper. Funding: Supported by NSF grants PHY-2012039 (D.S.W., Y.L. and Y.Z.), PHY-2012145 (Y.Z. and M.R.), and DMR-1653271 (S.G.). The computations were carried out at the Institute for Computational and Data Sciences at Penn State. Author contributions:  Y.L. carried out the experiments and experimental analysis; Y.Z. carried out the theoretical calculations; D.S.W. oversaw the experimental work; M.R. and D.S.W. oversaw the theoretical work. All authors were involved in the analysis and interpretation of the results, and all contributed to writing the paper.

\bibliographystyle{biblev1}
\bibliography{references}

\begin{thebibliography}{10}
\expandafter\ifx\csname url\endcsname\relax
  \def\url#1{{\tt #1}}\fi
\expandafter\ifx\csname urlprefix\endcsname\relax\def\urlprefix{URL }\fi
\expandafter\ifx\csname bibinfo\endcsname\relax\def\bibinfo#1#2{#2}\fi
\expandafter\ifx\csname eprint\endcsname\relax\def\eprint#1{\url{#1}}\fi

\bibitem{Rapp_2004}
\bibinfo{author}{R.~Rapp}, \bibinfo{title}{Theory highlights of quark matter
  2004},
  \bibinfo{journal}{\href{http://dx.doi.org/10.1088/0954-3899/30/8/038}{Journal
  of Physics G: Nuclear and Particle Physics}}
  \href{http://dx.doi.org/10.1088/0954-3899/30/8/038}{{\bf
  \bibinfo{volume}{30}}, \bibinfo{pages}{S951}}
  (\href{http://dx.doi.org/10.1088/0954-3899/30/8/038}{\bibinfo{year}{2004}}).

\bibitem{Florkowski_2018}
\bibinfo{author}{W.~Florkowski}, \bibinfo{author}{M.~P. Heller}, and
  \bibinfo{author}{M.~Spali{\'{n}}ski}, \bibinfo{title}{New theories of
  relativistic hydrodynamics in the {LHC} era},
  \bibinfo{journal}{\href{http://dx.doi.org/10.1088/1361-6633/aaa091}{Reports
  on Progress in Physics}}
  \href{http://dx.doi.org/10.1088/1361-6633/aaa091}{{\bf \bibinfo{volume}{81}},
  \bibinfo{pages}{046001}}
  (\href{http://dx.doi.org/10.1088/1361-6633/aaa091}{\bibinfo{year}{2018}}).

\bibitem{Schenke_2021}
\bibinfo{author}{B.~Schenke}, \bibinfo{title}{The smallest fluid on earth},
  \bibinfo{journal}{\href{http://dx.doi.org/10.1088/1361-6633/ac14c9}{Reports
  on Progress in Physics}}
  \href{http://dx.doi.org/10.1088/1361-6633/ac14c9}{{\bf \bibinfo{volume}{84}},
  \bibinfo{pages}{082301}}
  (\href{http://dx.doi.org/10.1088/1361-6633/ac14c9}{\bibinfo{year}{2021}}).

\bibitem{RevModPhys_2021}
\bibinfo{author}{J.~Berges}, \bibinfo{author}{M.~P. Heller},
  \bibinfo{author}{A.~Mazeliauskas}, and \bibinfo{author}{R.~Venugopalan},
  \bibinfo{title}{Qcd thermalization: Ab initio approaches and
  interdisciplinary connections},
  \bibinfo{journal}{\href{http://dx.doi.org/10.1103/RevModPhys.93.035003}{Rev.
  Mod. Phys.}} \href{http://dx.doi.org/10.1103/RevModPhys.93.035003}{{\bf
  \bibinfo{volume}{93}}, \bibinfo{pages}{035003}}
  (\href{http://dx.doi.org/10.1103/RevModPhys.93.035003}{\bibinfo{year}{2021}}).

\bibitem{prethermalization}
\bibinfo{author}{J.~Berges}, \bibinfo{author}{S.~Bors\'anyi}, and
  \bibinfo{author}{C.~Wetterich}, \bibinfo{title}{Prethermalization},
  \bibinfo{journal}{\href{http://dx.doi.org/10.1103/PhysRevLett.93.142002}{Phys.
  Rev. Lett.}} \href{http://dx.doi.org/10.1103/PhysRevLett.93.142002}{{\bf
  \bibinfo{volume}{93}}, \bibinfo{pages}{142002}}
  (\href{http://dx.doi.org/10.1103/PhysRevLett.93.142002}{\bibinfo{year}{2004}}).

\bibitem{Borsanyi}
\bibinfo{author}{S.~Borsanyi}, \bibinfo{title}{Prethermalization and the first
  1 fm/c of a heavy ion collision}, \bibinfo{journal}{Acta Physica Hungarica
  Series A, Heavy Ion Physics} {\bf \bibinfo{volume}{22}}, \bibinfo{pages}{317}
   (\bibinfo{year}{2005}).

\bibitem{rigol_dunjko_07}
\bibinfo{author}{M.~Rigol}, \bibinfo{author}{V.~Dunjko},
  \bibinfo{author}{V.~Yurovsky}, and \bibinfo{author}{M.~Olshanii},
  \bibinfo{title}{Relaxation in a completely integrable many-body quantum
  system: {A}n \textit{Ab Initio} study of the dynamics of the highly excited
  states of {1D} lattice hard-core bosons},
  \bibinfo{journal}{\href{http://dx.doi.org/10.1103/PhysRevLett.98.050405}{Phys.
  Rev. Lett.}} \href{http://dx.doi.org/10.1103/PhysRevLett.98.050405}{{\bf
  \bibinfo{volume}{98}}, \bibinfo{pages}{050405}}
  (\href{http://dx.doi.org/10.1103/PhysRevLett.98.050405}{\bibinfo{year}{2007}}).

\bibitem{schmiedmayer_prethermalization}
\bibinfo{author}{M.~Gring}, \bibinfo{author}{M.~Kuhnert},
  \bibinfo{author}{T.~Langen}, \bibinfo{author}{T.~Kitagawa},
  \bibinfo{author}{B.~Rauer}, \bibinfo{author}{M.~Schreitl},
  \bibinfo{author}{I.~Mazets}, \bibinfo{author}{D.~A. Smith},
  \bibinfo{author}{E.~Demler}, and \bibinfo{author}{J.~Schmiedmayer},
  \bibinfo{title}{Relaxation and prethermalization in an isolated quantum
  system},
  \bibinfo{journal}{\href{http://dx.doi.org/10.1126/science.1224953}{Science}}
  \href{http://dx.doi.org/10.1126/science.1224953}{{\bf \bibinfo{volume}{337}},
  \bibinfo{pages}{1318}}
  (\href{http://dx.doi.org/10.1126/science.1224953}{\bibinfo{year}{2012}}).

\bibitem{D_Alessio_2016}
\bibinfo{author}{L.~D’Alessio}, \bibinfo{author}{Y.~Kafri},
  \bibinfo{author}{A.~Polkovnikov}, and \bibinfo{author}{M.~Rigol},
  \bibinfo{title}{From quantum chaos and eigenstate thermalization to
  statistical mechanics and thermodynamics},
  \bibinfo{journal}{\href{http://dx.doi.org/10.1080/00018732.2016.1198134}{Adv.
  Phys.}} \href{http://dx.doi.org/10.1080/00018732.2016.1198134}{{\bf
  \bibinfo{volume}{65}}, \bibinfo{pages}{239–362}}
  (\href{http://dx.doi.org/10.1080/00018732.2016.1198134}{\bibinfo{year}{2016}}).

\bibitem{mallayya_rigol_19}
\bibinfo{author}{K.~Mallayya}, \bibinfo{author}{M.~Rigol}, and
  \bibinfo{author}{W.~De~Roeck}, \bibinfo{title}{Prethermalization and
  thermalization in isolated quantum systems},
  \bibinfo{journal}{\href{http://dx.doi.org/10.1103/PhysRevX.9.021027}{Phys.
  Rev. X}} \href{http://dx.doi.org/10.1103/PhysRevX.9.021027}{{\bf
  \bibinfo{volume}{9}}, \bibinfo{pages}{021027}}
  (\href{http://dx.doi.org/10.1103/PhysRevX.9.021027}{\bibinfo{year}{2019}}).

\bibitem{doyon2019lecture}
\bibinfo{author}{B.~Doyon}, \bibinfo{title}{Lecture notes on generalised
  hydrodynamics},
  \bibinfo{journal}{\href{http://dx.doi.org/10.21468/SciPostPhysLectNotes.18}{SciPost
  Phys. Lect. Notes}}
  \href{http://dx.doi.org/10.21468/SciPostPhysLectNotes.18}{\bibinfo{pages}{18}}
  (\href{http://dx.doi.org/10.21468/SciPostPhysLectNotes.18}{\bibinfo{year}{2020}}).

\bibitem{girardeau1960relationship}
\bibinfo{author}{M.~Girardeau}, \bibinfo{title}{Relationship between systems of
  impenetrable bosons and fermions in one dimension}, \bibinfo{journal}{Journal
  of Mathematical Physics} {\bf \bibinfo{volume}{1}}, \bibinfo{pages}{516}
  (\bibinfo{year}{1960}).

\bibitem{Spalinksy_2015}
\bibinfo{author}{M.~P. Heller} and
  \bibinfo{author}{M.~Spali\ifmmode~\acute{n}\else \'{n}\fi{}ski},
  \bibinfo{title}{Hydrodynamics beyond the gradient expansion: Resurgence and
  resummation},
  \bibinfo{journal}{\href{http://dx.doi.org/10.1103/PhysRevLett.115.072501}{Phys.
  Rev. Lett.}} \href{http://dx.doi.org/10.1103/PhysRevLett.115.072501}{{\bf
  \bibinfo{volume}{115}}, \bibinfo{pages}{072501}}
  (\href{http://dx.doi.org/10.1103/PhysRevLett.115.072501}{\bibinfo{year}{2015}}).

\bibitem{schlichting_2019}
\bibinfo{author}{G.~Giacalone}, \bibinfo{author}{A.~Mazeliauskas}, and
  \bibinfo{author}{S.~Schlichting}, \bibinfo{title}{Hydrodynamic attractors,
  initial state energy, and particle production in relativistic nuclear
  collisions},
  \bibinfo{journal}{\href{http://dx.doi.org/10.1103/PhysRevLett.123.262301}{Phys.
  Rev. Lett.}} \href{http://dx.doi.org/10.1103/PhysRevLett.123.262301}{{\bf
  \bibinfo{volume}{123}}, \bibinfo{pages}{262301}}
  (\href{http://dx.doi.org/10.1103/PhysRevLett.123.262301}{\bibinfo{year}{2019}}).

\bibitem{mottola_1997}
\bibinfo{author}{F.~Cooper}, \bibinfo{author}{S.~Habib},
  \bibinfo{author}{Y.~Kluger}, and \bibinfo{author}{E.~Mottola},
  \bibinfo{title}{Nonequilibrium dynamics of symmetry breaking in
  $\ensuremath{\lambda}{\ensuremath{\Phi}}^{4}$ theory},
  \bibinfo{journal}{\href{http://dx.doi.org/10.1103/PhysRevD.55.6471}{Phys.
  Rev. D}} \href{http://dx.doi.org/10.1103/PhysRevD.55.6471}{{\bf
  \bibinfo{volume}{55}}, \bibinfo{pages}{6471}}
  (\href{http://dx.doi.org/10.1103/PhysRevD.55.6471}{\bibinfo{year}{1997}}).

\bibitem{Kollar}
\bibinfo{author}{M.~Alexander} and \bibinfo{author}{M.~Kollar},
  \bibinfo{title}{Photoinduced prethermalization phenomena in correlated
  metals},
  \bibinfo{journal}{\href{http://dx.doi.org/https://doi.org/10.1002/pssb.202100280}{physica
  status solidi (b)}}
  \href{http://dx.doi.org/https://doi.org/10.1002/pssb.202100280}{{\bf
  \bibinfo{volume}{259}}, \bibinfo{pages}{2100280}}.

\bibitem{lieb1963exact}
\bibinfo{author}{E.~H. Lieb} and \bibinfo{author}{W.~Liniger},
  \bibinfo{title}{Exact analysis of an interacting {B}ose gas. {I. T}he general
  solution and the ground state},
  \bibinfo{journal}{\href{http://dx.doi.org/10.1103/PhysRev.130.1605}{Phys.
  Rev.}} \href{http://dx.doi.org/10.1103/PhysRev.130.1605}{{\bf
  \bibinfo{volume}{130}}, \bibinfo{pages}{1605}}
  (\href{http://dx.doi.org/10.1103/PhysRev.130.1605}{\bibinfo{year}{1963}}).

\bibitem{Cauxseptimescales_2016}
\bibinfo{author}{R.~van~den Berg}, \bibinfo{author}{B.~Wouters},
  \bibinfo{author}{S.~Eli\"ens}, \bibinfo{author}{J.~De~Nardis},
  \bibinfo{author}{R.~M. Konik}, and \bibinfo{author}{J.-S. Caux},
  \bibinfo{title}{Separation of time scales in a quantum {N}ewton's cradle},
  \bibinfo{journal}{\href{http://dx.doi.org/10.1103/PhysRevLett.116.225302}{Phys.
  Rev. Lett.}} \href{http://dx.doi.org/10.1103/PhysRevLett.116.225302}{{\bf
  \bibinfo{volume}{116}}, \bibinfo{pages}{225302}}
  (\href{http://dx.doi.org/10.1103/PhysRevLett.116.225302}{\bibinfo{year}{2016}}).

\bibitem{minguzzi_gangardt_05}
\bibinfo{author}{A.~Minguzzi} and \bibinfo{author}{D.~M. Gangardt},
  \bibinfo{title}{Exact coherent states of a harmonically confined
  {Tonks-Girardeau} gas},
  \bibinfo{journal}{\href{http://dx.doi.org/10.1103/PhysRevLett.94.240404}{Phys.
  Rev. Lett.}} \href{http://dx.doi.org/10.1103/PhysRevLett.94.240404}{{\bf
  \bibinfo{volume}{94}}, \bibinfo{pages}{240404}}
  (\href{http://dx.doi.org/10.1103/PhysRevLett.94.240404}{\bibinfo{year}{2005}}).

\bibitem{del2022hydrodynamic}
\bibinfo{author}{G.~D.~V. Del~Vecchio} and \bibinfo{author}{B.~Doyon},
  \bibinfo{title}{The hydrodynamic theory of dynamical correlation functions in
  the xx chain}, \bibinfo{journal}{Journal of Statistical Mechanics: Theory and
  Experiment} {\bf \bibinfo{volume}{2022}}, \bibinfo{pages}{053102}
  (\bibinfo{year}{2022}).

\bibitem{wilson_malvania_20}
\bibinfo{author}{J.~M. Wilson}, \bibinfo{author}{N.~Malvania},
  \bibinfo{author}{Y.~Le}, \bibinfo{author}{Y.~Zhang},
  \bibinfo{author}{M.~Rigol}, and \bibinfo{author}{D.~S. Weiss},
  \bibinfo{title}{Observation of dynamical fermionization},
  \bibinfo{journal}{\href{http://dx.doi.org/10.1126/science.aaz0242}{Science}}
  \href{http://dx.doi.org/10.1126/science.aaz0242}{{\bf \bibinfo{volume}{367}},
  \bibinfo{pages}{1461}}
  (\href{http://dx.doi.org/10.1126/science.aaz0242}{\bibinfo{year}{2020}}).

\bibitem{Malvania_2021}
\bibinfo{author}{N.~Malvania}, \bibinfo{author}{Y.~Zhang},
  \bibinfo{author}{Y.~Le}, \bibinfo{author}{J.~Dubail},
  \bibinfo{author}{M.~Rigol}, and \bibinfo{author}{D.~S. Weiss},
  \bibinfo{title}{Generalized hydrodynamics in strongly interacting 1d bose
  gases},
  \bibinfo{journal}{\href{http://dx.doi.org/10.1126/science.abf0147}{Science}}
  \href{http://dx.doi.org/10.1126/science.abf0147}{{\bf \bibinfo{volume}{373}},
  \bibinfo{pages}{1129}}
  (\href{http://dx.doi.org/10.1126/science.abf0147}{\bibinfo{year}{2021}}).

\bibitem{castro2016emergent}
\bibinfo{author}{O.~A. Castro-Alvaredo}, \bibinfo{author}{B.~Doyon}, and
  \bibinfo{author}{T.~Yoshimura}, \bibinfo{title}{Emergent hydrodynamics in
  integrable quantum systems out of equilibrium},
  \bibinfo{journal}{\href{http://dx.doi.org/10.1103/PhysRevX.6.041065}{Phys.
  Rev. X}} \href{http://dx.doi.org/10.1103/PhysRevX.6.041065}{{\bf
  \bibinfo{volume}{6}}, \bibinfo{pages}{041065}}
  (\href{http://dx.doi.org/10.1103/PhysRevX.6.041065}{\bibinfo{year}{2016}}).

\bibitem{bertini2016transport}
\bibinfo{author}{B.~Bertini}, \bibinfo{author}{M.~Collura},
  \bibinfo{author}{J.~De~Nardis}, and \bibinfo{author}{M.~Fagotti},
  \bibinfo{title}{Transport in out-of-equilibrium {XXZ} chains: {E}xact
  profiles of charges and currents},
  \bibinfo{journal}{\href{http://dx.doi.org/10.1103/PhysRevLett.117.207201}{Phys.
  Rev. Lett.}} \href{http://dx.doi.org/10.1103/PhysRevLett.117.207201}{{\bf
  \bibinfo{volume}{117}}, \bibinfo{pages}{207201}}
  (\href{http://dx.doi.org/10.1103/PhysRevLett.117.207201}{\bibinfo{year}{2016}}).

\bibitem{schemmer2019generalized}
\bibinfo{author}{M.~Schemmer}, \bibinfo{author}{I.~Bouchoule},
  \bibinfo{author}{B.~Doyon}, and \bibinfo{author}{J.~Dubail},
  \bibinfo{title}{Generalized hydrodynamics on an atom chip},
  \bibinfo{journal}{\href{http://dx.doi.org/10.1103/PhysRevLett.122.090601}{Phys.
  Rev. Lett.}} \href{http://dx.doi.org/10.1103/PhysRevLett.122.090601}{{\bf
  \bibinfo{volume}{122}}, \bibinfo{pages}{090601}}
  (\href{http://dx.doi.org/10.1103/PhysRevLett.122.090601}{\bibinfo{year}{2019}}).

\bibitem{kinoshita_all_optical}
\bibinfo{author}{T.~Kinoshita}, \bibinfo{author}{T.~Wenger}, and
  \bibinfo{author}{D.~S. Weiss}, \bibinfo{title}{All-optical bose-einstein
  condensation using a compressible crossed dipole trap},
  \bibinfo{journal}{\href{http://dx.doi.org/10.1103/PhysRevA.71.011602}{Phys.
  Rev. A}} \href{http://dx.doi.org/10.1103/PhysRevA.71.011602}{{\bf
  \bibinfo{volume}{71}}, \bibinfo{pages}{011602}}
  (\href{http://dx.doi.org/10.1103/PhysRevA.71.011602}{\bibinfo{year}{2005}}).

\bibitem{olshanii1998atomic}
\bibinfo{author}{M.~Olshanii}, \bibinfo{title}{Atomic scattering in the
  presence of an external confinement and a gas of impenetrable bosons},
  \bibinfo{journal}{\href{http://dx.doi.org/10.1103/PhysRevLett.81.938}{Phys.
  Rev. Lett.}} \href{http://dx.doi.org/10.1103/PhysRevLett.81.938}{{\bf
  \bibinfo{volume}{81}}, \bibinfo{pages}{938}}
  (\href{http://dx.doi.org/10.1103/PhysRevLett.81.938}{\bibinfo{year}{1998}}).

\bibitem{yang1969thermodynamics}
\bibinfo{author}{C.~N. Yang} and \bibinfo{author}{C.~P. Yang},
  \bibinfo{title}{Thermodynamics of a one‐dimensional system of bosons with
  repulsive delta‐function interaction},
  \bibinfo{journal}{\href{http://dx.doi.org/10.1063/1.1664947}{J. Math. Phys.}}
  \href{http://dx.doi.org/10.1063/1.1664947}{{\bf \bibinfo{volume}{10}},
  \bibinfo{pages}{1115}}
  (\href{http://dx.doi.org/10.1063/1.1664947}{\bibinfo{year}{1969}}).

\bibitem{tolra_ohara_04}
\bibinfo{author}{B.~L. Tolra}, \bibinfo{author}{K.~M. O'Hara},
  \bibinfo{author}{J.~H. Huckans}, \bibinfo{author}{W.~D. Phillips},
  \bibinfo{author}{S.~L. Rolston}, and \bibinfo{author}{J.~V. Porto},
  \bibinfo{title}{Observation of reduced three-body recombination in a
  correlated 1d degenerate bose gas},
  \bibinfo{journal}{\href{http://dx.doi.org/10.1103/PhysRevLett.92.190401}{Phys.
  Rev. Lett.}} \href{http://dx.doi.org/10.1103/PhysRevLett.92.190401}{{\bf
  \bibinfo{volume}{92}}, \bibinfo{pages}{190401}}
  (\href{http://dx.doi.org/10.1103/PhysRevLett.92.190401}{\bibinfo{year}{2004}}).

\bibitem{rigol_muramatsu_05}
\bibinfo{author}{M.~Rigol} and \bibinfo{author}{A.~Muramatsu},
  \bibinfo{title}{Ground-state properties of hard-core bosons confined on
  one-dimensional optical lattices},
  \bibinfo{journal}{\href{http://dx.doi.org/10.1103/PhysRevA.72.013604}{Phys.
  Rev. A}} \href{http://dx.doi.org/10.1103/PhysRevA.72.013604}{{\bf
  \bibinfo{volume}{72}}, \bibinfo{pages}{013604}}
  (\href{http://dx.doi.org/10.1103/PhysRevA.72.013604}{\bibinfo{year}{2005}}).

\bibitem{xu_rigol_17}
\bibinfo{author}{W.~Xu} and \bibinfo{author}{M.~Rigol},
  \bibinfo{title}{Expansion of one-dimensional lattice hard-core bosons at
  finite temperature},
  \bibinfo{journal}{\href{http://dx.doi.org/10.1103/PhysRevA.95.033617}{Phys.
  Rev. A}} \href{http://dx.doi.org/10.1103/PhysRevA.95.033617}{{\bf
  \bibinfo{volume}{95}}, \bibinfo{pages}{033617}}
  (\href{http://dx.doi.org/10.1103/PhysRevA.95.033617}{\bibinfo{year}{2017}}).

\end{thebibliography}

\newpage

%
%

\vspace{1cm}
\begin{center}
{\bf \large Methods}
\end{center}

\section{Experimental setup} \label{model}
We create a BEC of $^{87}$Rb in the $F = 1$, $m_{f} = 1$ ground state by evaporative cooling in a compressible red-detuned crossed dipole trap made with 1064 nm wavelength light and final beam waists of 57 $\mu$m~\cite{kinoshita_all_optical}. We control the atom number in the BEC, which ranges from $2\times10^5$ to $3.4\times10^5$ atoms, by varying the final evaporation depth of the crossed dipole trap. We then ramp up the crossed dipole trap to powers that range from 5.8-157 mW, which allows us to start with a range of initial trap densities. We next create a bundle of 1D gases by adiabatically ramping up a blue-detuned 40$E_r$-deep 2D lattice made with 432 $\mu$m diameter, 139 mW, 775 nm wavelength, retroreflected pairs of crossed beams. For our Bragg scattering experiment, we pulse on a similar axial lattice beam pair. The lattice and Bragg beam pairs are all mutually offset from each other by rf frequencies to prevent mutual interference. The depth of the Bragg lattice is 27$E_r$. During the 6~$\mu$s pulse, the incipient first momentum sidebands move $\sim70$~nm, which is not negligibly small compared to the lattice period of 387.5~nm. The evolution of the system during the pulse slightly modifies the shape and relative peak heights of the rapidity distributions compared to a shorter Kapitza-Dirac pulse on a non-interacting gas, but the difference is qualitatively unimportant to the post-pulse evolution that we study in this paper. 

\section{Lieb-Liniger model}
The experimental system can be modeled as a 2D array of independent 1D gases. Each 1D gas is described by the LL Hamiltonian in the presence of a confining potential $U(z)$ \cite{lieb1963exact},
\begin{equation}\label{H_lieb_liniger}
{\cal H}_\text{LL}=\sum_{j=1}^{N}\left[-\frac{\hbar^2}{2m}\frac{\partial^2}{\partial z^2_j}+U(z_j)\right]+g\sum_{1\leq j < l \leq N}\delta(z_j-z_l) \,,
\end{equation}
where $m$ is the mass of a $^{87}$Rb atom, and $N$ is the number of atoms. $g$($>0$ in our case) is the strength of effective 1D contact interaction, which depends on the depth of 2D lattice \cite{olshanii1998atomic}. In the absence of confining potential $U(z)$, Hamiltonian (\ref{H_lieb_liniger}) is exactly solvable via the Bethe ansatz \cite{lieb1963exact, yang1969thermodynamics}. All observables in the equilibrium states depend only on the dimensionless coupling strength $\gamma=mg/n_{\rm 1D}\hbar^2$, where $n_{\rm 1D}$ is the particle density in the 1D gas. We use the local $\gamma(z)=mg/n_{\rm 1D}(z)\hbar^2$ in the trapped system.

The number of particles in each 1D gas depends on its $(x,y)$ position, which can be modeled as
\begin{equation}\label{TF_distribution}
N(x,y)=N(0,0)\bigg[1-\frac{x^2+y^2}{R_{\rm TF}^2}\bigg]^{3/2}\,,
\end{equation}
where $N(0,0)$ is the number of particles in the central 1D gas and $R_{\rm TF}$ is the Thomas-Fermi radius~\cite{tolra_ohara_04}. The total number of atom in the system is $N_{\rm tot}=\sum_{x,y}N(x,y)$. With the experimentally measured $R_{\rm TF}$ and $N_{\rm tot}$, we can find $N(x,y)$. The averaged $\bar\gamma$ for the experimental system is defined as,
\begin{equation}\label{gamma_avg}
\bar\gamma=\frac{1}{N_{\rm tot}}\sum_{x,y}\int n_{\rm 1D}(z; x,y)\gamma(z;x,y)dz\,.
\end{equation}
$n_{\rm 1D}(z;x,y)$ can be calculated from the Bethe ansatz solution of homogeneous LL model with a local density approximation (LDA), with the knowledge of $N(x,y)$ and $U(z)$. 

\section{Momentum and rapidity measurements}

To measure momentum distributions, we turn off the 2D lattice and the crossed dipole trap suddenly and let the atoms expand in free space for a time of flight, TOF (TOF = 42~ms for $\bar{\gamma} = 2.3-7.6$ and TOF = 45~ms for $\bar{\gamma_0} = 0.94$), before taking an absorption image~\cite{wilson_malvania_20}. When the 2D lattice is turned off, the atom cloud expands rapidly in the transverse direction and the energies associated with atom-atom interactions decrease significantly before the momentum distribution can evolve. The initial cloud sizes range from 15 to 25~$\mu$m for $\bar{\gamma}_0=0.94$ to 7.6. The TOF is limited by the transverse expansion that is essential to the momentum measurement (see Ref. \cite{wilson_malvania_20}).
  
To measure rapidity distributions, we expand the atoms in an approximately flat potential in 1D until the distribution reaches its asymptotic shape~\cite{wilson_malvania_20, Malvania_2021}. The flat potential is made by leaving a shallow axial trap on to cancel the anti-trap from the blue-detuned lattice. After the Bragg pulse, the rapidity distribution of the central peak does not change on the short time scale of $\sim0.5$ ms, before the movement of atoms in the axial trap starts to change the distribution (see Extended Data Fig.~2). The flat potential has a finite size of $\sim40\,\mu$m, which means that the rapidity measurement is only accurate for the central peak. The side peaks are distorted, since they fall down the potential hill of the blue-detuned lattice. However, the distortion does not significantly change during the first $\sim0.5$ ms of evolution, so the message of Extended Data Fig.~4, that the rapidities do not evolve in this time, is not undermined. We only show rapidities measurements for the condition $\bar{\gamma}_0 = 2.3$, because the flat potential is not long enough when $\bar{\gamma}_0$ is higher and the side peaks are not well enough separated from the central peak when $\bar{\gamma}_0$ is lower.

\section{Energy Error bar}
Each time point in the energy curves of Fig.~\ref{hydrodynamization} is obtained from an average of 10 images, taken in two groups of 5 during the experiment. To minimize extra noise associated with background drifts, it is necessary to average at least 5 images before extracting the energy. We combine the data for these averages in two ways, first by combining the images within each temporally separated group of 5 and second by combining them into groups of odd and even images. For each grouping approach, we calculate average errors by taking the average over all time points of the root mean square difference between the two groups. The grouping makes no difference for the $\bar\gamma_0=2.3$ data, but for the other data the average error tends to be higher for the temporally separated groups, which is indicative of small experimental drifts. The average errors calculated using temporally separated groups are higher than the ones calculated using the odd and even group by 130$\%$ for the $\bar\gamma_0=0.94$ data and 70$\%$ for the $\bar\gamma_0=3.4$ data. To account for the error associated with the drift, without overstating its contribution to the total error, for all the curves we use the average of the average errors obtained with the two grouping approaches.

\section{Time constants from the $p_f$ curves}

The shape of the $p_f(t_{ev})$ curves are not universal (see Fig.~\ref{prethermalization}), so they will not all fit to the same simple function. There are two reasons for the differences among these curves. First, there are things that affect the calculated or measured momentum distributions. Theoretically, momentum distributions for different densities are affected differently by finite number effects. Experimentally, high $\bar{\gamma}_0$ momentum distributions are somewhat broadened by finite-size contributions to the momentum measurements. They are completely negligible for $\bar{\gamma}_0=0.94$ and become more significant at higher $\bar{\gamma}_0$, where the initial cloud lengths are larger (as large as 25 $\mu$m for $\bar{\gamma}_0=6.9$) and the kinetic energies are smaller. As a check on the importance of this effect, we deconvolved the initial size effects under the assumption that the initial momentum distributions are the theoretical bosonic distributions. We find that the extracted time constants barely change as a result, so we present only the directly measured $p_f(t_{ev})$s.

The second reason that these curves have different shapes relates to the motion of the side peaks. The $p_f(t_{ev})$ curves eventually decay from their peak value because the local GGE changes as the side peaks progressively overlap less with the central peak. The effect is larger and faster when the density is higher and the initial cloud length is smaller, i.e., at lower $\bar{\gamma}_0$ or higher average density. The effect is also evident from the difference between the $\bar{\gamma}_0=3.4$ and 3.7 curves in Fig.~\ref{prethermalization}a, where the difference in $\bar{\gamma}_0$ is due to a difference in initial density. 

We have tried two ways to extract relatively universal time constants from these varied shape curves. For the first method, which we use in Fig.~\ref{scaling}, we determined the minimum and maximum $p_{50}$ for the initial rise and find the average of the two, (${p_{50}}_{\rm h} = ({p_{50}}_{\rm max}+{p_{50}}_{\rm min})/2$). We determine the time at which ${p_{50}}_{\rm h}$ is reached by fitting the three nearest to ${p_{50}}_{\rm h}$ points to a straight line. The second method was to identify how long it takes to rise from 20$\%$ to 80$\%$ of the maximum change in $p_{50}$, where those points were determined with local linear fits. Both approaches give similar results for the time constant as a function of 1/$\sqrt{\epsilon}$ (see Fig.~\ref{scaling}), with quantitatively comparable linear dependence. We chose to use the first method because it gives somewhat more similar values for the $\bar{\gamma}_0=3.4$ and 3.7 curves, despite their rather different shapes. For $\bar{\gamma}_0$ = 0.94, 2.3, and 3.4, the $p_f(t_{ev})$ curves were taken with about three times as many points as at the earliest times. To minimize the effect of fluctuations on the time constants, we smooth the curves by averaging each $p_{f}(t_{ev})$ with two neighboring points. The smoothed curves are then used to extract the time constants.

To calculate the error bars for the time constants, we first find $\Delta{p_{50}}_{\rm h}$ by combining in quadrature the errors from determining the minimum and maximum of the $p_{50}$ ($\Delta {p_{50}}_{\rm min}$, $\Delta {p_{50}}_{\rm max}$) with the fitting error associated with the offset from ${p_{50}}_{\rm h}$ of the linear fit of the three points near ${p_{50}}_{\rm h}$. We then use the slope of the linear fit line to convert $\Delta{p_{50}}_{\rm h}$ to $\Delta\tau_{p_{50}}$.

The theoretical time constants plotted in Fig.~\ref{scaling} are extracted as follows. We first smooth the theoretical curve $\tilde p_{50}(t_{ev})$ by averaging each $p_{50}(t_{ev})$ with its neighboring points within a time interval $\delta t_{ev}$, $\tilde p_{50}(t_{ev}) =\sum_{t\in[t_{ev}\pm\delta t_{ev}/2]}p_{50}(t)/N_{\delta t_{ev}}$. $N_{\delta t_{ev}}$ is the total number of data points used for the average. By choosing $\delta t_{ev}$ such that it matches the oscillation period for hydrodynamization, we try to minimize its effect on the local prethermalization time constant. Then, as when extracting the experimental time constants, we define the time constant as the time where $\tilde p_{50_{\rm h}}=(\tilde p_{50_{\rm max}}+\tilde p_{50_{\rm min}})/2$ is reached. We estimate the error associated with the smoothing procedure to be $\Delta p_{50_{\rm h}}=\sqrt{\sum_{t}[p_{50}(t)-\tilde p_{50}(t)]^2/(N_{\delta t_{ev}}N_{t})}$. The average is done for $t\in[t_{30\%},t_{70\%}]$ where $t_{30\%}$ ($t_{70\%}$) is the time when $\tilde p_{50}$ changes by 30\% (70\%).

\section{Numerical simulations} \label{sec:numerical}

All the numerical calculations are done in the TG limit ($\gamma\to\infty$) of the LL Hamiltonian (\ref{H_lieb_liniger}). Specifically, we use the low-site-occupancy regime of the lattice hard-core boson Hamiltonian,
\begin{equation}\label{H_HCB}
{\cal H}_{\rm HCB}=-J\sum_{j=1}^{L-1}\big(\hat b^\dagger_{j+1}\hat b^{}_j + {\rm H.c.}\big)+\sum_{j=1}^{L}U(z_j)\hat b^\dagger_j\hat b^{}_j\,,
\end{equation}
where $J$ is the hopping amplitude and $L$ is the total number of lattice sites. $\hat b^\dagger_{j}$ ($\hat b^{}_j$) creates (annihilates) a hard-core boson at site $j$, with the additional constraints $\hat {b}^{\dagger2}_j=\hat b^{\,2}_j=0$ to enforce the hard-core condition. The position of site $j$ in the lattice is taken to be $z_j=(j-L/2)a$, where $a$ is the lattice spacing. In the limit of vanishing site occupancy ($n_j=\langle\hat {b}^{\dagger}_j\hat {b}^{}_j\rangle\to0$ at all sites), in which the average distance between particles is much larger than $a$, the lattice Hamiltonian (\ref{H_HCB}) is equivalent to the continuum TG limit of the LL Hamiltonian (\ref{H_lieb_liniger})~\cite{rigol_muramatsu_05, wilson_malvania_20}. The parameters of the two Hamiltonians satisfy the relation $J=\hbar^2/(2ma^2)$. 

The one-body correlation functions of the 1D lattice hard-core bosons are computed exactly by mapping the hard-core boson Hamiltonian onto noninteracting spinless fermions via the Jordan-Wigner transformation, and then using properties of Slater determinants \cite{rigol_muramatsu_05, xu_rigol_17}. For the ground state calculations, we choose the lattice spacing to be $a=5\times 10^{-9}$~m and simulate systems with up to $L=12000$ lattice sites. We verify that lattice effects are negligible by doing some of the calculations on a lattice with twice as large $a$ and checking that the results do not change within the desired accuracy. Due to the much higher computational cost of the finite-temperature calculations~\cite{xu_rigol_17}, the finite-temperature data presented in Extended Data Fig.~8 was obtained for $a=3.2\times 10^{-8}$~m ($a=4\times 10^{-8}$~m) on a lattice with $L=1500$ sites for $T=5$~nK (10~nK).

Our numerical calculations start at $t=0$ from the ground (or finite temperature) state of Eq.~(\ref{H_HCB}) with $N$ particles and a Gaussian shaped trapping potential
\begin{equation}\label{U_ini}
U(z)=U_0\left[1-\exp\left(-\frac{2z^2}{W^2}\right)\right]\,,
\end{equation}
where $U_0$ is the strength of the Gaussian trap and $W$ is the trap width. As in the experiments, the quench is implemented by evolving the initial state under Hamiltonian~\eqref{H_HCB} with the addition of the Bragg pulse potential (characterized by an amplitude $U_{\rm pulse}$ and a wave number $k$) 
\begin{equation}\label{U_pulse}
U_{pulse}(z)=U_{pulse}\cos^2(kz)\,,
\end{equation}
for a time $t_{pulse}$. At times $t=t_{pulse}+t_{ev}$, the system evolves under the initial Hamiltonian~(\ref{U_ini}), and it is during those times that we calculate the momentum and rapidity distributions. For Fig.~\ref{hydrodynamization} and Extended Data Figs.~1, 3, 4, and 8, we use $t_{pulse}=6~\mu$s, as in the experiment. Figs.~\ref{prethermalization}, \ref{scaling}, and Extended Data Figs.~7c and 7d all use $k=4k_0$ in order to shorten the hydrodynamization time scale relative to local prethermalization time scales. To minimize many-body evolution during the Bragg pulse, we decrease $t_{pulse}$ to 1 $\mu$s in those calculations, while increasing $U_{pulse}$ in order to keep the fraction of atoms in the central peak fixed.

%
%

\newpage
\begin{center}
{\bf \large Extended Data}
\end{center}

\setcounter{figure}{0}
\renewcommand{\figurename}{{\bf Extended Data Figure}}
\linespread{1}

\begin{figure}[!h]
\includegraphics[width=0.95\columnwidth]{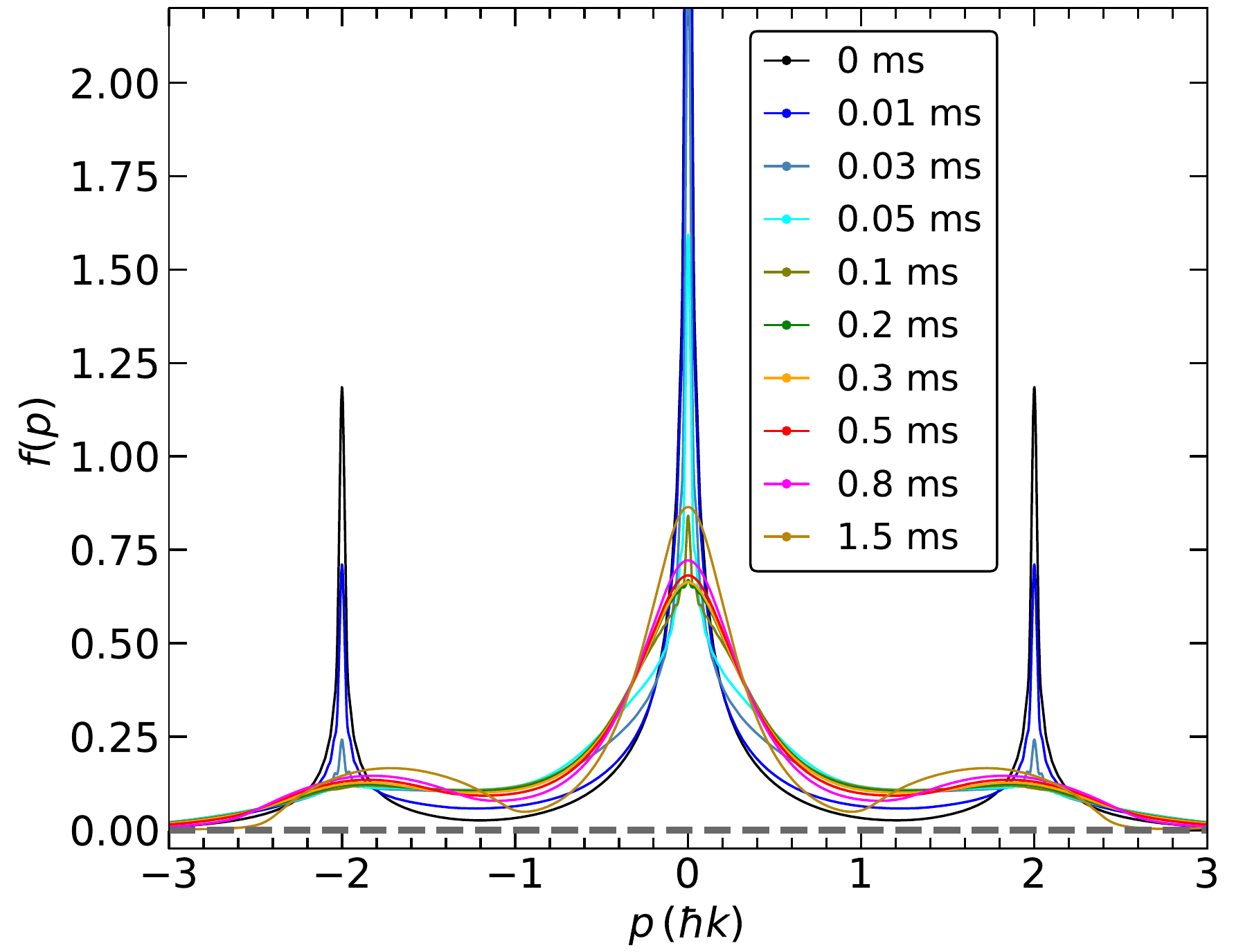}
\caption{{\bf Theoretical momentum distributions.} Calculations are done for a single 1D tube in the TG limit at zero temperature. We use the same trap and quench parameters as in the $\bar\gamma_0=2.3$ experiment, and choose the atom number in the tube to be $N=32$ in order to match the experimental average energy density. These curves match the experimental evolution times in Fig.~1.}
\label{MomentumDist_theory}
\end{figure}

\begin{figure}[!t]
\includegraphics[width=0.95\columnwidth]{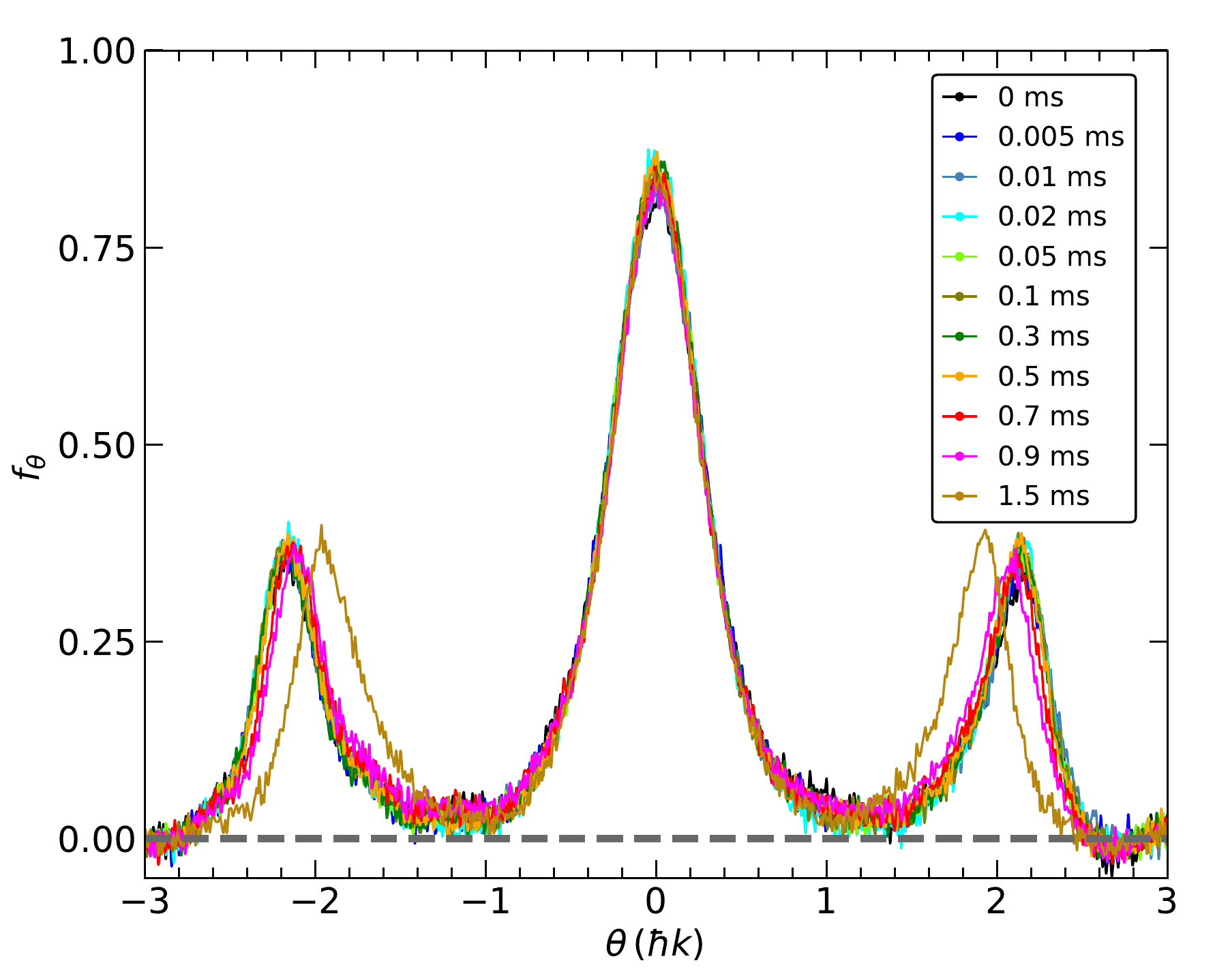}
\caption{{\bf Experimental rapidity distributions} for $\bar \gamma_0 = 2.3$ at a sequence of times after the Bragg scattering quench. The central part of the rapidity distribution does not appreciably change over 1.5 ms, which is $<10\%$ of the trap oscillation period. The rapidities measurement for the side peaks are slightly distorted by the fact that the flat potential does not extend far enough for atoms moving that fast (after they have moved $\sim$20$\mu$m, they are accelerated slightly). Still, since the distortion is approximately the same for all times within the first $\sim$0.5~ms, the fact that the measured distribution does not change implies that that part of the rapidity distribution also remains constant. After $\sim$1~ms, the side peaks start to be noticeably slowed as they climb up the potential of the Gaussian axial trap.}
\label{Rapidity_exp}
\end{figure}

\begin{figure}[!t]
\includegraphics[width=0.95\columnwidth]{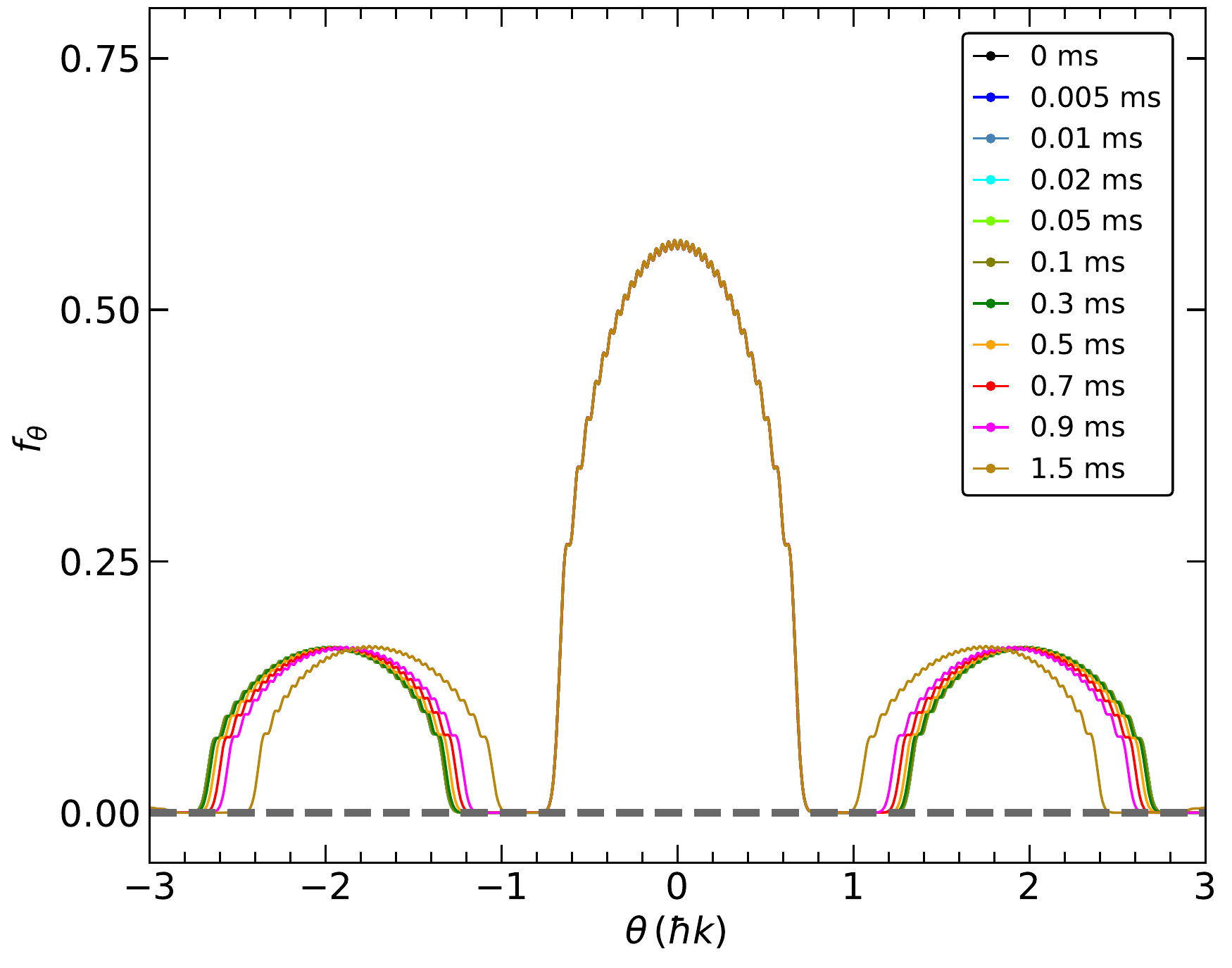}
\caption{{\bf Theoretical rapidity distributions.} These simulations are done for a single 1D tube in the TG limit at zero temperature with same trap and quench parameters as in the $\bar\gamma_0=2.3$ experiment. The atom number in the tube is chosen to be $N=32$ in order to match the experimental average energy density. These curves match the experimental evolution times of Extended Data Fig.~\ref{Rapidity_exp}.}
\label{Rapidity_theory}
\end{figure}

\begin{figure}[!t]
\includegraphics[width=0.62\columnwidth]{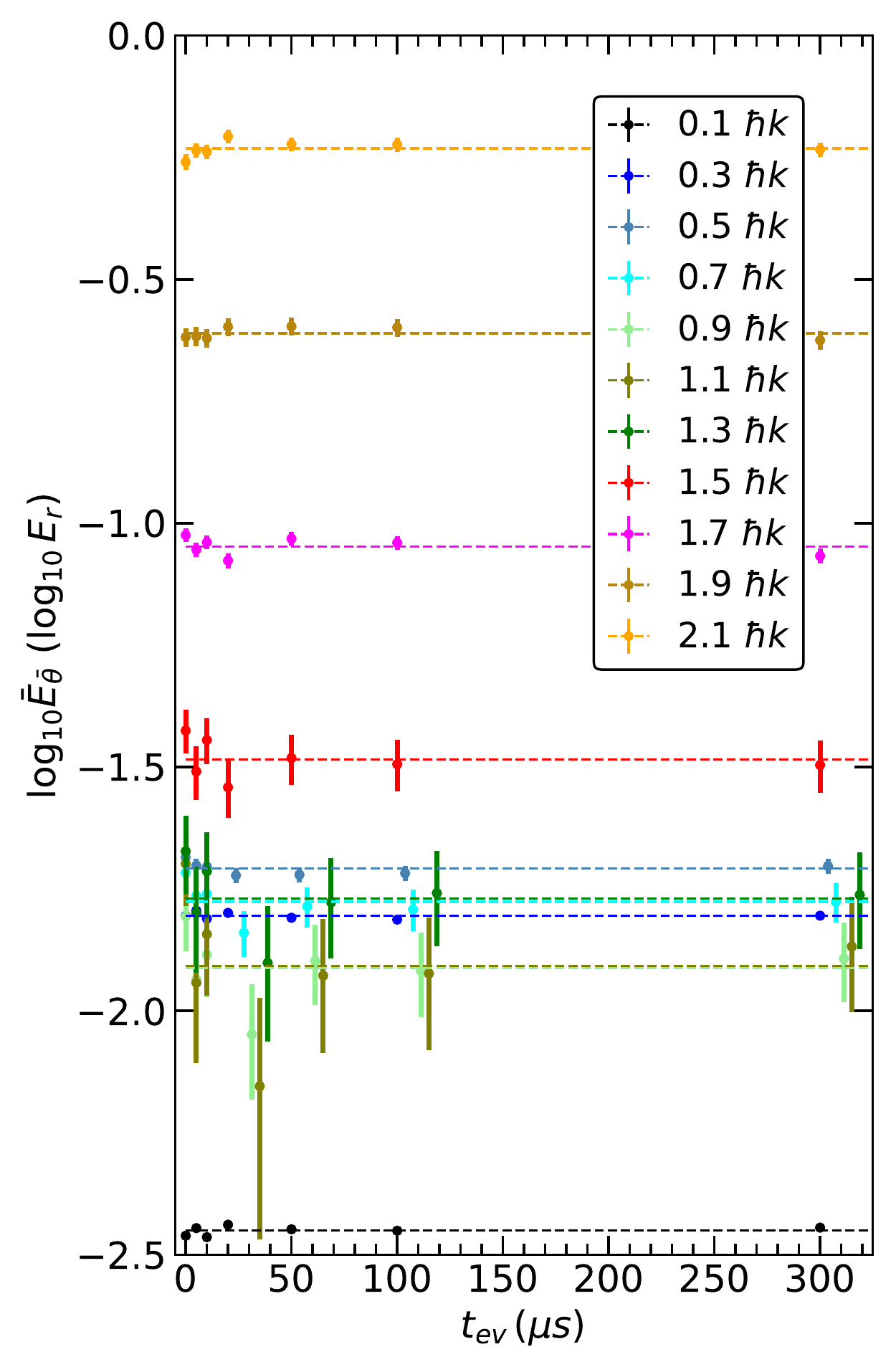}
\caption{{\bf Integrated experimental rapidity energy}. The curves are extracted from the rapidity profiles for $\bar \gamma_0 = 2.3$ integrating over different 0.2 $\hbar k$ wide rapidity groups. The different colors denote different rapidity groups (as in Fig.~2), defined in the key. For the average rapidity $\bar \theta = 0.5 \hbar k - 1.3 \hbar k$ curves, we have horizontally shifted the points for times longer than $t_{ev} = 20~\mu$s in order to better resolve the different rapidity group energies at the same time. The dashed lines show the average energy for each rapidity group. There is no detectable change in the energy of each rapidity group in the first $\sim300~\mu$s,  in stark contrast to the 3 dB-scale rapid changes in the energies associated with momentum groups in Fig.~2b.}
\label{RapidityEnergy_exp}
\end{figure}

\begin{figure}[!t]
\includegraphics[width=0.58\columnwidth]{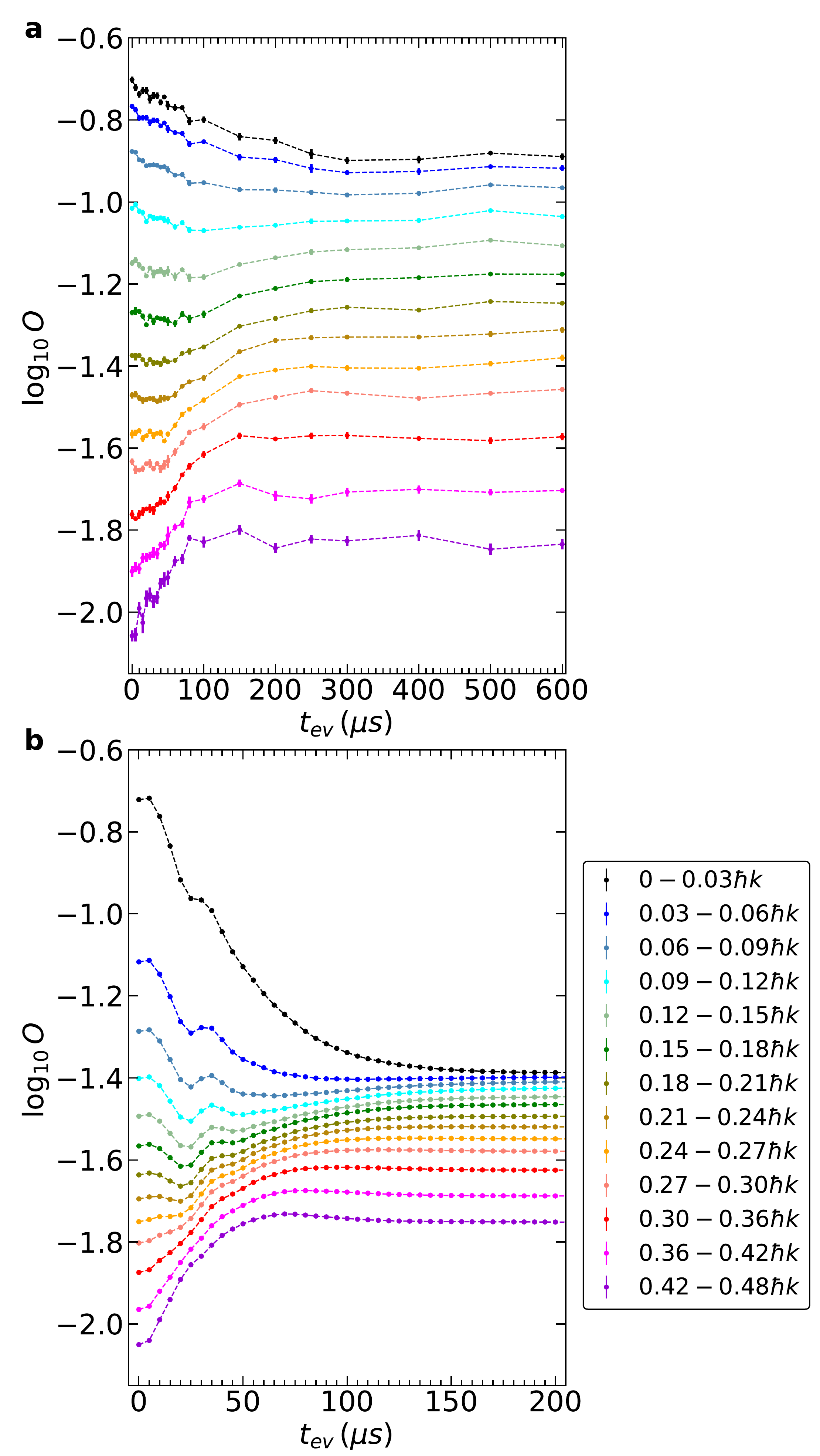}
\caption{{\bf The time evolution of the occupations of different momentum groups within the central peak.} \textbf{a}. Experimental curves for $\bar \gamma_0 = 2.3$ are plotted on a log scale. Each curve is obtained by integrating the area of the normalized momentum distribution within the designated momentum range. Different colors denote the different momentum groups as shown in the legend. The last three momentum groups, $0.30-0.36 \hbar k$, $0.36-0.42 \hbar k$, and $0.42-0.48 \hbar k$, have twice the integration range, so their occupations are divided by 2. \textbf{b}. Theoretical curves for a single 1D gas in the TG limit with the same average energy as in the experiment with $\bar \gamma_0 = 2.3$. In both the experiment and the theory, the occupation of higher momentum groups evolves faster than the lower ones, as expected for local prethermalization. The fact that each of these curves has a different shape makes it difficult to quantitatively compare time constants among them. The theoretical curves evolve more than twice as fast as the experimental curves, presumably reflecting the difference between infinite and finite $g$ (see Eq.~1). $k=k_0$ for the experiment and $k=4k_0$ for the theory. }
\label{MomentumOccupancy}
\end{figure}

\begin{figure}[!t]
\includegraphics[width=0.69\columnwidth]{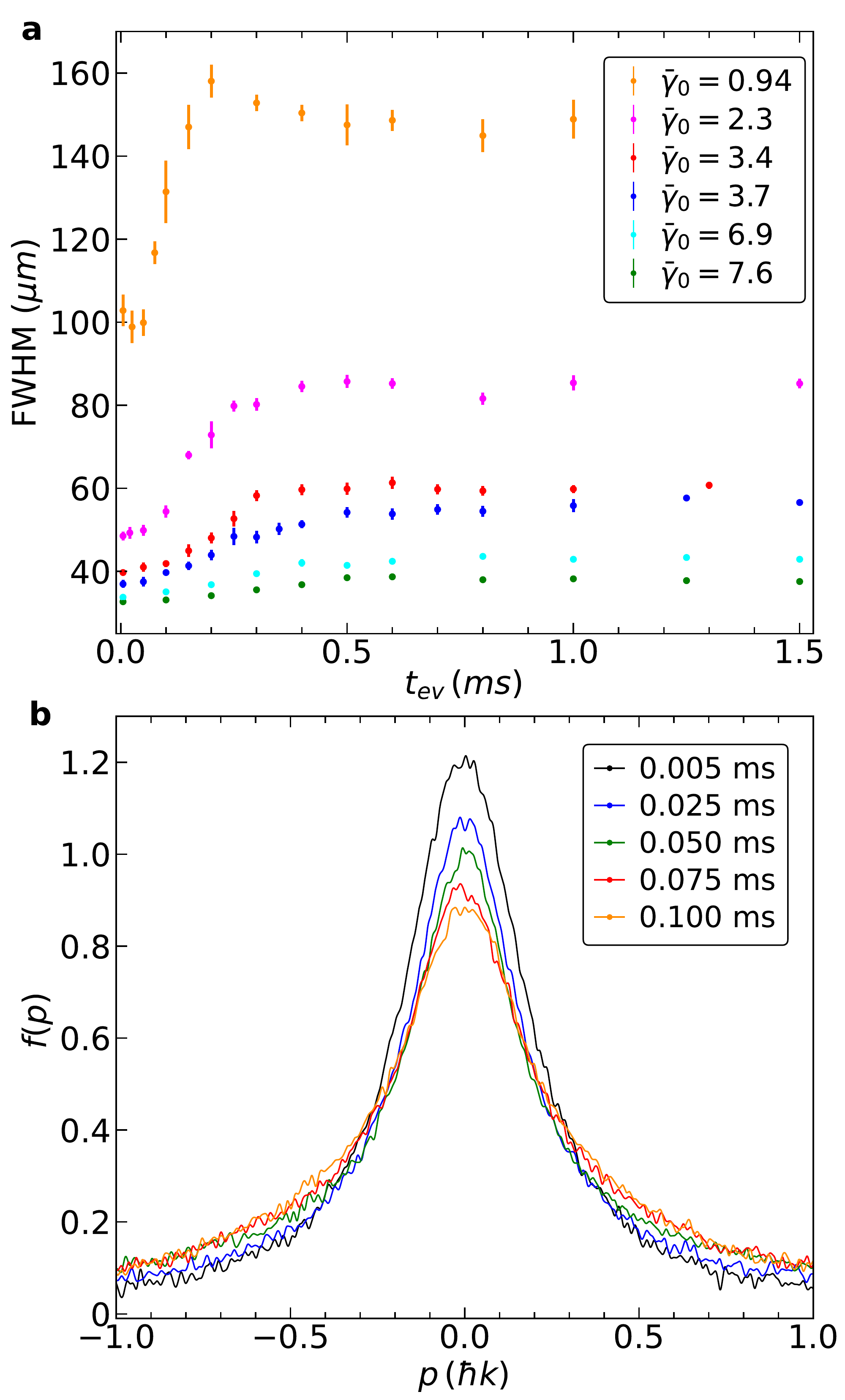}
\caption{{\bf Time evolution of the FWHM.} \textbf{a}. The evolution of the FWHM of the central peak in the momentum distributions after the Bragg scattering quench for different coupling strengths. To a greater degree than for the $p_{50}(t_{ev})$ curves of Fig.~3, these curves all have different shapes. \textbf{b}. The evolving momentum distribution of the central peak for $\bar{\gamma}_0 = 0.94$. These curves correspond to the first five points of the orange curve in \textbf{a}. The momentum distribution clearly evolves during the first 0.05~ms, even though the FWHM does not change.  This illustrates that the FWHM is not a reliable marker of the evolution of these momentum distributions.}
\label{FWHM_exp}
\end{figure}

\begin{figure}[!t]
\includegraphics[width=0.65\columnwidth]{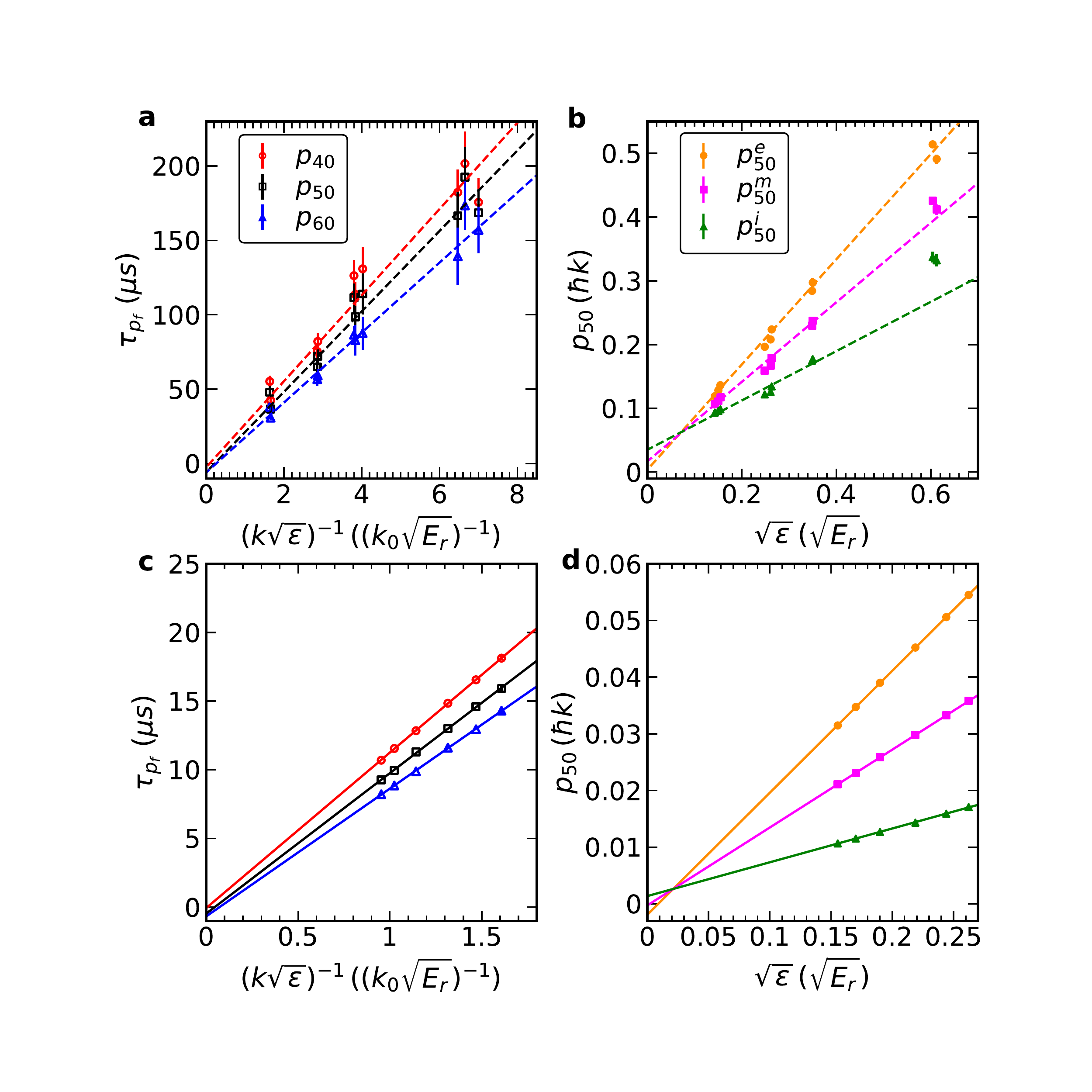}
\caption{{\bf Relationships among $p_{f}$, $\tau_{p_f}$, and $\epsilon$.} \textbf{a}. Experimental time constants associated with $p_{40}$ (red circles), $p_{50}$ (black squares), and $p_{60}$ (blue triangles) as functions of $1/k\sqrt{\epsilon}$. As in all the experiments presented in this paper, $k$ is fixed at $k_0$. The time constants are extracted from curves like those in Fig.~3 (see Methods). The dashed lines are least-squares linear fits; the intercepts are $-2.5 \pm 5.6$, $-5.8 \pm 4.9$, $-6.0 \pm 3.5$ for $p_{40}$, $p_{50}$, and $p_{60}$, respectively. The data is consistent with linear relationships between the time constant associated with each $p_f$ and the inverse of the square root of average energy per particle. For a given momentum distribution, the actual values of $p_{40}$ and $p_{60}$ span a range of $\sim\pm$30\% around the steepest part of the distribution (at $\sim p_{50}$). \textbf{b}. Experimental $p_{50}$ vs $\sqrt{\epsilon}$. The green triangles, magenta squares, and orange circles correspond to the initial ($p_{50}^{i}$), middle ($p_{50}^{m}$), and final ($p_{50}^{e}$) values of $p_{50}$ for each experimental condition, extracted from Fig.~3. The dashed lines are least-squares linear fits; the intercepts are $(3.5 \pm 0.9)\times10^{-2}$, $(1.7 \pm 0.6)\times10^{-2}$, and $(3.8 \pm 4.4)\times10^{-3}$ for $p_{50}^{i}$, $p_{50}^{m}$, and $p_{50}^{e}$, respectively. The $p_{50}^{i}$ points at the lowest $\epsilon$ (highest $\bar \gamma_0$) conditions are more likely to be affected by finite-size corrections to their momentum distributions. The data shows a linear relationship between each measured value and $\sqrt{\epsilon}$. \textbf{c}. Theoretical time constants obtained from the $p_{40}$, $p_{50}$, and $p_{60}$ curves simulated with $k = 4k_0$. The time constants are obtained from curves like those in the inset in Fig.~3. The error bars are smaller than the marker size. The solid lines are least-squares linear fits; the intercepts are $-0.065 \pm 0.047$, $-0.47 \pm 0.13$, and $-0.67 \pm 0.10$ for $p_{40}$, $p_{50}$, and $p_{60}$, respectively. \textbf{d}. Theoretical $p_{50}$ vs $\sqrt{\epsilon}$, simulated with $k = 4k_0$. The intercepts are ($14 \pm 1.3)\times10^{-4}$, $(2.7\pm 0.53)\times10^{-4}$, and $(-19 \pm 1.2)\times10^{-4}$, for $p_{50}^{i}$, $p_{50}^{m}$, and $p_{50}^{e}$, respectively. The momentum feature that is most clearly proportional to $\sqrt{\epsilon}$ is $p_{50}^{m}$. Since that is the midpoint value of $p_{50}$ during the evolution, it is likely to be close to $\bar{p}_{50}$, the effective momentum to which the $p_{50}$ measurement is sensitive. Taken all together, this figure shows that, for $p_{50}$, the time constants are proportional to $1/\sqrt{\epsilon}$, which is in turn proportional to the characteristic momentum being measured. We have repeated the entire analysis for $p_{40}$ and $p_{60}$ and the conclusions are the same.}
\label{p40p60p50}
\end{figure}

\begin{figure}[!t]
\includegraphics[width=0.95\columnwidth]{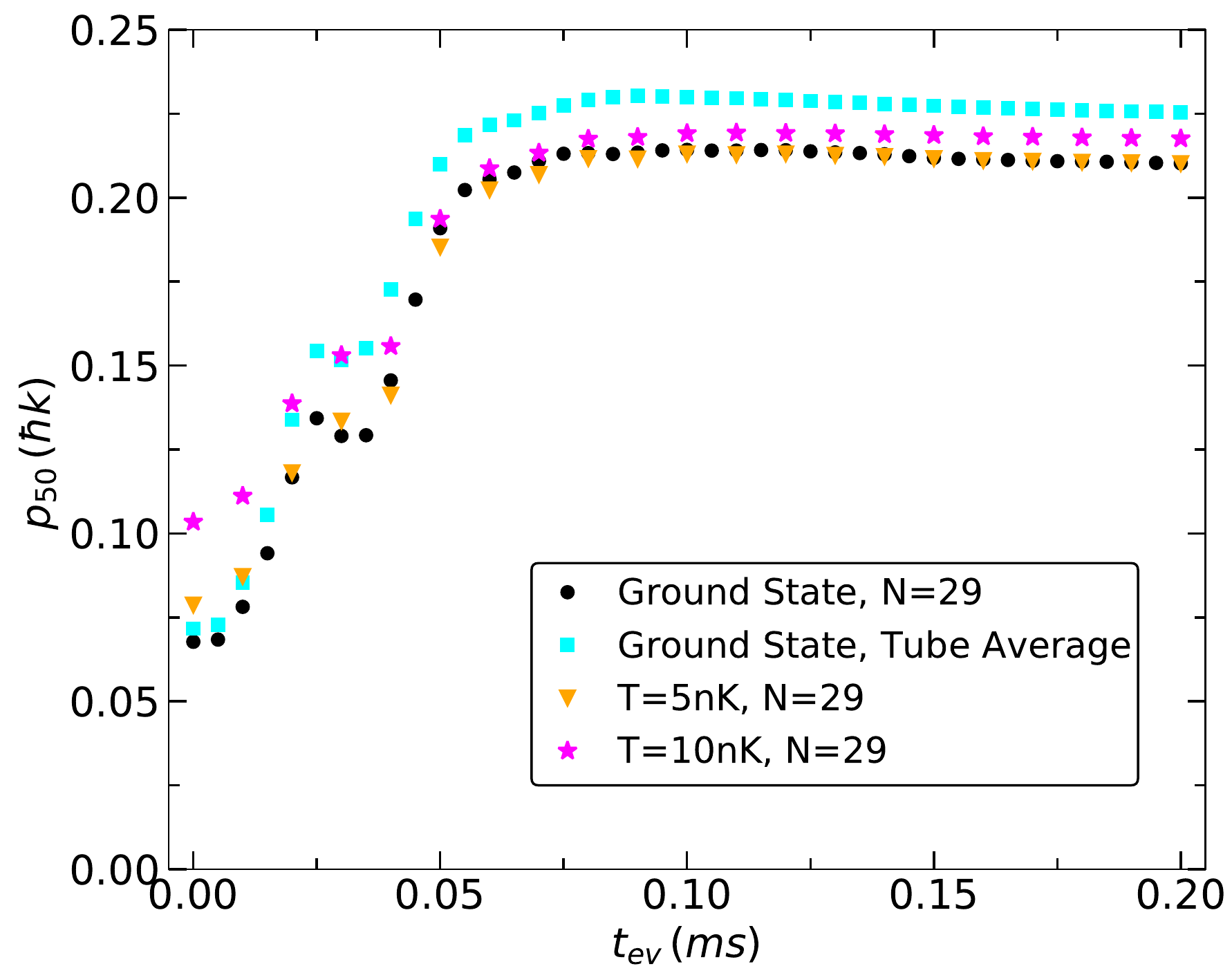}
\caption{{\bf Effect of the average over 1D gases and of finite temperature on $p_{50}$ .} We plot theoretical simulations of $p_{50}$ in the TG limit with the same trap and quench parameters as in the $\bar\gamma_0=3.4$ experiment. For the average over 1D gases, we use the Thomas-Fermi distribution with experimentally measured $R_{\rm TF}=23\mu m$ and a total particle number of $N_{\rm tot}=2.75\times10^5$ (see Methods). To simplify the calculations, we round the particle number in each tube in steps of 5. The circles show the results of ground state simulations for a single tube with $N=29$ particles (which matches the experimental average energy density). The squares show the results of ground state simulations after averaging over all the 1D gases, as occurs in the experimental setup. The triangles (stars) show simulations for a single tube with $N=29$ particles at a temperature of $T=5$ nK (10nK). We did the finite temperature simulations with a larger discretization $a=3.2\times10^{-8}$ m ($a=4\times10^{-8}$ m) due to numerical limitations (see Methods). The results show no significant changes in the time constant due to either the average over tubes or finite temperature.}
\label{p50_theory}
\end{figure}

\bibliographystyleM{biblev1}
\bibliographyM{references}
\end{document}